\documentclass[longauth]{aa} 
\usepackage{graphicx}
\usepackage{txfonts}
\usepackage{array}
\usepackage{balance}
\PassOptionsToPackage{hyphens}{url}\usepackage[breaklinks]{hyperref}
\usepackage{natbib}

\newcommand{\Teff}{\ensuremath{T_\mathrm{eff}}}

\newcommand{\loggf}{\ensuremath{\log\,gf}}
\newcommand{\logg}{\ensuremath{\log\,g}}
\def\teff{$T\rm_{eff}$}
\newcommand{\kms}{$\rm km s ^{-1}$}

\newcommand{\ncstar}{HE\,0107$-$5240}

\begin{document} 

\title{ Unveiling the nature of \ncstar: a long period binary CEMP-no star with  [Fe/H] of --5.56
\thanks{Based on ESPRESSO GTO observations collected under ESO programmes   0103.D-0700 and  108.2268.001 (PI: P. Molaro), 1104.C-0350 1102.C-0744 (PI  F.  Pepe), and 114.27NH.001 (PI: D. Aguado)}
}
\titlerunning{\ncstar}

\author{
E.~Caffau  \inst{1,2} \and
M.~Steffen \inst{3} \and
P.~Molaro \inst{2,4} \and
P.~Bonifacio \inst{1,2} \and
N.~Christlieb \inst{5} \and
D.\,S.~Aguado \inst{6,7}\and
J.\,I.~Gonz\'alez Hern\'andez \inst{6,7} \and
M.\,R.~Zapatero Osorio \inst{8} \and
L.~Monaco \inst{9,2} \and
M.~Limongi \inst{10,11,12} \and
A.~Chieffi \inst{13,11,14} \and
A.~Falla \inst{15,11} \and
L.~Roberti \inst{16,11,17,18} \and
A.\,J.~Gallagher \inst{19} \and
M.~Spite \inst{1} \and
P.~Fran\c{c}ois \inst{1,20} \and
H.-G.~Ludwig \inst{5}\and
L.~Sbordone \inst{21} \and      
R.~Lallement \inst{1} \and
C.~Allende Prieto \inst{6,7} \and
R.~Rebolo \inst{6,7} \and
S.~Cristiani \inst{2,4} \and
G.~Cupani \inst{2,4} \and
V.~D'Odorico \inst{2,4} \and
C.\,J.\,A.\,P.~Martins \inst{22,23} \and
D.~Milakovi\'{c} \inst{2,4} \and
M.\,T.~Murphy \inst{24} \and
N.\,J.~Nunes \inst{25} \and
N.\,C.~Santos \inst{22,26} \and
T.\,M.~Schmidt \inst{27}
}
\institute{LIRA, Observatoire de Paris, Universit{\'e} PSL, Sorbonne Universit{\'e}, Universit{\'e} Paris Cit{\'e}, CY Cergy Paris Universit{\'e}, CNRS,92190 Meudon, France
\and
INAF-OATS, Via G.B.Tiepolo 11, Trieste, I 34143, Italy
\and
Leibniz-Institut f\"ur Astrophysik (AIP), An der Sternwarte 16, 14482 Potsdam, Germany
\and
Institute for Fundamental Physics of the Universe (IFPU), Via Beirut 2, I34151 Grignano, Trieste, Italy.
\and
Zentrum f\"ur Astronomie der Universit\"at Heidelberg, Landessternwarte,
K\"onigstuhl 12, 69117 Heidelberg, Germany
\and
Instituto de Astrofísica de Canarias, V\'ia L\'actea, 38205 La Laguna, Tenerife, Spain
\and
Universidad de La Laguna, Departamento de Astrof\'isica, 38206 La Laguna, Tenerife, Spain
\and
Centro de Astrobiolog\'ia (CSIC-INTA), Carretera Ajalvir km 4, 28850 Torrejón de Ardoz, Madrid, Spain
\and
Universidad Andres Bello, Facultad de Ciencias Exactas, Departamento de F\'isica y Astronom\'ia – Instituto de Astrof\'isica, Autopista Concepci\'on-Talcahuano 7100, Talcahuano, Chile
\and
Istituto Nazionale di Astrofisica - Osservatorio Astronomico di Roma (INAF - OAR), Via Frascati 33, I-00040, Monteporzio Catone, Italy
\and
Istituto Nazionale di Fisica Nucleare - Sezione di Perugia (INFN), via A. Pascoli s/n, I-06125 Perugia, Italy
\and
Kavli Institute for the Physics and Mathematics of the Universe, Todai Institutes for Advanced Study, University of Tokyo, Kashiwa, 277-8583 (Kavli IPMU, WPI), Japan
\and
Istituto Nazionale di Astrofisica - Istituto di Astrofisica e Planetologia Spaziali (INAF - IAPS), Via Fosso del Cavaliere 100, I-00133, Roma, Italy
\and
Monash Centre for Astrophysics (MoCA), School of Mathematical Sciences, Monash University, Victoria 3800, Australia
\and
Dipartimento di Fisica, Sapienza Università di Roma, P.le A. Moro 5, Roma 00185, Italy
\and
Istituto Nazionale di Fisica Nucleare - Laboratori Nazionali del Sud (INFN - LNS), Via Santa Sofia 62, Catania, Italy
\and
Konkoly Observatory, Research Centre for Astronomy and Earth Sciences, E\"otv\"os Lor\'and Research Network (ELKH), Konkoly Thege Mikl\'{o}s \'{u}t 15-17, H-1121 Budapest, Hungary
\and
CSFK, MTA Centre of Excellence, Konkoly Thege Miklós út 15-17, H-1121, Budapest, Hungary
\and
Goethe University Frankfurt, Institute for Applied Physics (IAP), Max-von-Laue-Str. 12, 60438 Frankfurt am Main, Germany
\and
UPJV, Universit\'e de Picardie Jules Verne, 33 rue St Leu, 80080 Amiens, France
\and
European Southern Observatory, Casilla 19001, Santiago, Chile
\and
 Instituto de Astrof\'{i}sica e Ci\^{e}ncias do Espa\c{c}o, Universidade do Porto, CAUP, Rua das Estrelas, 4150-762 Porto, Portugal 
\and
Centro de Astrof\'{\i}sica da Universidade do Porto, Rua das Estrelas, 4150-762 Porto, Portugal
\and
Centre for Astrophysics and Supercomputing, Swinburne University of Technology, Hawthorn, Victoria 3122, Australia
\and
Instituto de Astrof\'isica e Ci\^{e}ncias do Espa\c{c}o, Faculdade de Ci\^encias da Universidade de Lisboa,
Campo Grande, PT1749-016 Lisboa, Portugal
\and
Departamento de F\'{i}sica e Astronomia, Faculdade de Ci\^{e}ncias, Universidade do Porto, Rua do Campo Alegre, 4169-007 Porto, Portugal
 \and
Observatoire Astronomique de l’Universit\'e de Gen\`eve, Chemin Pegasi 51, Sauverny, CH-1290, Switzerland
}

   \date{Received July 15, 2024; accepted August 16, 2024}
   
  \abstract
{The vast majority of the most iron-poor stars in the Galaxy exhibit a strong carbon enhancement, with C/H ratios only about two orders of magnitude below solar. 
This unusual chemical composition likely reflects the properties of the gas cloud from which these stars formed, having been enriched by one, or at most a few, supernovae.
A remarkable member of this stellar class, \ncstar\ with [Fe/H]=--5.56, has been identified as part of a binary system.}
{To constrain its orbital parameters, radial velocity monitoring has been carried out using the ESPRESSO spectrograph.}
{Radial velocities were derived using cross-correlation with a template, taking advantage of the strong G-band feature. Combining all observations yielded a high signal-to-noise spectrum, which has been used to refine our understanding of the stellar chemical composition. Additionally, a co-added UVES spectrum in the blue was used to complement the wavelength coverage of ESPRESSO.}
{Observations of \ncstar\ over a span of more than four years have yielded a revised orbital period 
of about 29 years.
Updated elemental abundances have been determined for Sc, Cr, Co, and, tentatively, Al, along with a new upper limit for Be. The iron abundance has been derived from  ionised Fe lines. Significant upper limits have been established for Li, Si, and Sr. 
}
{The star is confirmed to be a long-period binary.
Iron abundances derived from  neutral and ionised lines are consistent with local thermodynamical equilibrium (LTE) assumption, casting doubt on published deviation from LTE corrections for Fe for this star. The heavy elements Sr and Ba remain undetected, confirming the classification of \ncstar\ as a carbon enhanced metal-poor and non enhanced in heavy elements (CEMP-no) star and supporting the absence of an n-capture element plateau at the lowest metallicities.}

\keywords{Stars: abundances - Galaxy: abundances - Galaxy: evolution - Galaxy: formation}
 \maketitle
 
\nolinenumbers
\section{Introduction\label{intro}}

The metal-poor stars shining today in the sky are expected to have formed shortly after the Big Bang from gas clouds that were only slightly enriched by one or a few supernova explosions. Given their life times comparable to the age of the Universe, they must have low masses ($\rm M < 0.8M_\odot.$)
Metal-poor stars have been known since the early work of \citet{chamberlain51}, but it was the HK survey by \citet{beers1992} that led to the discovery of a distinct group: carbon-enhanced, iron-poor stars. Within this group, a subclass was identified that combines high carbon enhancement with a deficiency or absence of n-capture elements \citep{norris1997,bonifacio1998}. The star \ncstar\ was discovered at the beginning of this century \citep{christlieb02} and has since become the prototype of this rare class, which now includes most of the most iron-poor stars known.

Currently, about a dozen stars are known with extremely low iron abundances ($\rm [Fe/H] < -4.5$) and relatively high carbon abundances ($\rm 5.5 < A(C) < 7.8$). Only two known stars in this regime (SDSS\,J102915.14+172927.9 \citealt{leo11} and Pristine\,J221.8781+09.7844 \citealt{starkenburg18}) do not exhibit carbon enhancement, having only upper limits on their carbon abundances: $\rm A(C) < 4.68$ for the former \citep{leo2024} and below 6.0 for the latter \citep{PristineXIV}.
Recently \citet{limberg2025}, reported of an evolved star which shows no C in the spectrum and, after taking into account for the lost C into N, the upper limit is just $\rm A(C)<4.79$ ($\rm [C/Fe]<+1.18$).

Un-evolved, warm stars (\teff$>5800$\,K) often display a weak G-band, making carbon abundance determinations challenging and complicating the identification of extremely metal poor  stars, EMP,  that are not carbon-enhanced (i.e., not CEMP, a CEMP being here defined as $\rm [C/Fe] > 1.0$, \citealt{beers05}). According to \citet{topos2}, CEMP-no stars show a relatively constant absolute carbon abundance, lower than that of CEMP-s stars, which also show enhanced abundances of s-process elements. The carbon in CEMP-s stars likely originates from mass transfer from an evolved asymptotic giant branch (AGB) companion, along with elements such as Sr and Ba. In contrast, the carbon in CEMP-no stars is likely primordial. The so-called `low-carbon band' \citep{topos2} is centred around $\rm A(C) \sim 6.8$. We therefore suggest that in the ultra-iron-poor regime ($\rm [Fe/H] < -4.5$), stars with $\rm A(C) > 5.8$ should be classified as CEMP, rather than relying solely on $\rm [C/Fe] > 1$.

It is plausible that carbon-normal and carbon-enhanced ultra iron-poor stars have distinct origins. For example, SDSS\,J102915.14+172927.9 exhibits a solar-scaled abundance pattern without the typical $\alpha$-element enhancement seen in most metal-poor stars and an orbit in the Galactic disc. 
This may indicate that this is a Pop\,III star and 
that its stellar surface was polluted during its long journey through the Galactic disc \citep[see][]{leo2024}.

\ncstar\ was first reported by \citet{christlieb02}, with a detailed chemical abundance analysis presented in \citet{christlieb04}, based on UVES spectra. The oxygen abundance was later derived from OH molecular lines in the UV by \citet{bessel04}. The impact of 3D hydrodynamic models on the inferred abundances was explored by \citet{collet06}, while deviations from local thermodynamic equilibrium (NLTE) were examined for iron by \citet{ezzeddine17} and for Mg and Ca by \citet{sitnova19}.

The star \ncstar\ was included in an ESO ESPRESSO GTO programme aimed at investigating binarity among the most metal-poor stars. Its binary nature was first suggested by \citet{arentsen19}, who noted discrepancies between radial velocities measured with UVES and SALT. This was subsequently confirmed by \citet{bonifacio20} using early ESPRESSO data. Continued ESPRESSO monitoring over more than three years was analysed by \citet{aguado22}, who derived a long orbital period of approximately 13,000 days ($\sim$36 years), showing that binary formation is not precluded at extremely low metallicities.
\citet{aguado22} also measured a carbon isotopic ratio of $^{12}$C/$^{13}$C = $87 \pm 6$. This value rules out mass transfer from an evolved AGB companion as the source of the observed carbon and implies a significant primordial production of $^{13}$C in the progenitor. In the present study, we extend the temporal baseline of radial velocity monitoring and analyse a new, high signal-to-noise co-added spectrum derived from all the available ESPRESSO observations.

\section{Observations} 
\ncstar\  was observed as part of several ESPRESSO GTO programmes: 0103.D-0700, 108.2268.001, and 110.245W.001 (PI: P. Molaro); 1102.C-0744 and 1104.C-0350 (PI: F. Pepe); as well as during the open-time programme 114.27NH.001 (PI: D. S. Aguado). The observations were reduced all in the same way. The GTO data, previously analysed by \citet{aguado22} and \citet{molaro23}, are here complemented with the additional open-time observations. All spectra cover the same wavelength range from 380 to 788\,nm, and were corrected for the star’s orbital motion, co-added, and used for a revised chemical abundance analysis.

Since the ESPRESSO spectrum extends only down to approximately 380\,nm in the blue, UVES data were employed to study chemical species with transitions in the ultraviolet. The UVES spectra consist of a sequence of observations using the 346\,nm setting, obtained between September 30 and December 27, 2002, for a total exposure time of 19.5 hours \citep[see][for the observation log]{bonifacio21}. A total of 22 observing blocks (OBs) from programme 70.D-0009 were available, of which 21 were usable (one OB was exposed for only 1\,s and contained no signal). Most OBs had exposure times of 3600\,s, with three exceptions at 1727, 1600, and 3960\,s, respectively. The combined UVES spectrum reached a signal-to-noise ratio (S/N) of approximately 50 per pixel at 324\,nm.

The individual UVES spectra were shifted to the rest frame, scaled relative to one another, and co-added using an iterative $\kappa\sigma$-clipping procedure to compute a weighted average while removing outliers. Due to the pixel scale of 0.215'' per pixel in the blue arm of UVES, the resulting spectrum was oversampled; it was therefore rebinned by a factor of two. The final co-added spectrum achieves a peak S/N per 0.05\,\AA\ pixel of $\sim 140$ at 3800\,\AA, decreasing to $\rm S/N\sim 70$ at 3350\,\AA\ and $\sim 30$ at 3100\,\AA.

The HRS at SALT spectrum \citep{arentsen19} has been used only for the radial velocity determination.

\begin{table}
\begin{center}
\caption{\label{tab:vel} Journal of the ESPRESSO observations with a measurement of the radial velocities.}
\scriptsize
\hspace*{0cm}
\begin{tabular}{lcccc}
 \hline
 Date & hour & MJD\tablefootmark{a} & $v_{rad}$        & error    \\
   &         &.                     & (km$\,$s$^{-1}$) & (km$\,$s$^{-1}$)   \\
 \hline
2018-09-03 &02:35:15.830  & 58364.1078 &48.170    &0.047   \\
 2019-08-03 &07:31:54.524  & 58698.3138 &47.925    &0.014   \\
 2019-08-26 &06:18:47.382  & 58721.2630 &47.873    &0.015   \\
 2019-09-04 &04:57:33.171  & 58741.2506 &47.905    &0.016   \\
 2019-10-03 &02:18:47.960  &58759.0963  &47.913     &0.015   \\
 2020-12-07 &02:19:17.802  &59190.0967  &47.741     &0.017   \\
 2021-01-23 &00:47:12.076  &59435.1762 &47.627    &0.016   \\
 2021-08-09 &05:16:11.004  &59435.2195 &47.647    &0.013   \\
 2021-10-02 &05:42:52.690  &59489.2381 &47.559    &0.022   \\
 2021-10-31 &01:28:39.476  &59518.0615 &47.593    &0.014   \\
 2021-11-11 &04:51:34.041  &59529.2024 &47.610    &0.012   \\
 2022-01-27 &01:00:20.859  &59606.0419 &47.525    &0.020   \\
 2022-10-05 &02:23:36.234  &59857.0997 &47.429    &0.014   \\
 2022-10-06 &05:15:07.307  &59858.2188 &47.408    &0.014   \\
 2022-10-07 &00:55:29.027  &59859.0385 &47.429    &0.019   \\
 2022-10-09 &01:55:47.567  &59861.0804 &47.446    &0.017   \\
 2022-10-10 &01:32:41.459  &59862.0643 &47.406    &0.013   \\
 2022-10-13 &06:35:05.896  &59865.2743 &47.496    &0.018   \\
 2022-10-14 &05:09:08.436  &59866.2146 &47.378    &0.012   \\
 2022-10-15 &06:16:20.113  &59867.2613 &47.383    &0.013   \\
 2022-10-16 &00:28:15.002  &59868.0196 &47.362    &0.011   \\
 2022-10-17 &01:28:38.559  &59869.0615 &47.411    &0.018   \\
 2022-10-17 &05:20:01.708  &59869.2222 &47.382    &0.013   \\
 2022-10-30 &00:36:16.959  &59882.0251 &47.391    &0.015   \\
 2022-11-04 &03:40:39.767  &59887.1532 &47.386    &0.014   \\
 2022-11-10 &03:37:47.954  &59893.1512 &47.437    &0.012   \\
 2022-11-17 &02:14:41.449  &59900.0935 &47.346    &0.016   \\
 2022-11-22 &02:34:23.004  &59905.1072 &47.395    &0.012   \\
 2022-11-27 &01:35:49.040  &59910.0665 &47.316    &0.015   \\
 2022-12-02 &01:18:39.614  &59915.0546 &47.350    &0.020   \\
 2022-12-03 &02:40:27.679  &59916.1114 &47.339    &0.016   \\
 2022-12-04 &01:01:41.885  &59917.0428 &47.381    &0.012   \\
 2023-06-09 &08:45:19.968  &60104.3648 &47.311    &0.012   \\
 2024-10-04 &01:12:31.436  &60587.0503 &46.841    &0.017   \\
 2024-11-03 &02:12:59.673  &60617.0923 &46.842    &0.011   \\
 2024-12-03 &03:02:13.415  &60647.1265 &46.857    &0.010   \\
\hline
\end{tabular}
\tablefoot{\tablefoottext{a}{Modified Julian date at the start of observation.}}
\end{center}
\end{table}

\section{Galactic Kinematics} \label{sec:kine}

We adopted coordinates, proper motions and parallax from Gaia DR3 \citep{gaiadr3}, applying the parallax zero-point correction as prescribed by \citet[][]{lindegren21}.
The distance derived from the parallax is of $6.93^{+1.1}_{-0.8}$\,kpc.
The star is part of a binary system, although the faint companion is not directly visible. Radial velocity variations have been extensively discussed by \citet{aguado22} and \citet{molaro23}. For our analysis, we adopted the systemic (barycentric) radial velocity of the system, $\rm V_{\rm rad} = 46.6 \pm 0.1$\,\kms\ (see Sec.\,\ref{sec:anal}), and used this value to derive the star’s Galactic orbit.

Following the procedure described in \citet{ghs143}, we employed the galpy code \citep{bovy15} with the MWPotential2014 Galactic potential model. We integrated the orbit backward in time over 1 Gyr. For the solar motion, we adopted the parameters from \citet[][]{schonrich10}, assuming a solar distance to the Galactic centre of 8\, kpc and a circular velocity of 220\,\kms\
\citep[][]{bovy12}.

\begin{figure}
\centering
\includegraphics[width=0.9\hsize,clip=true]{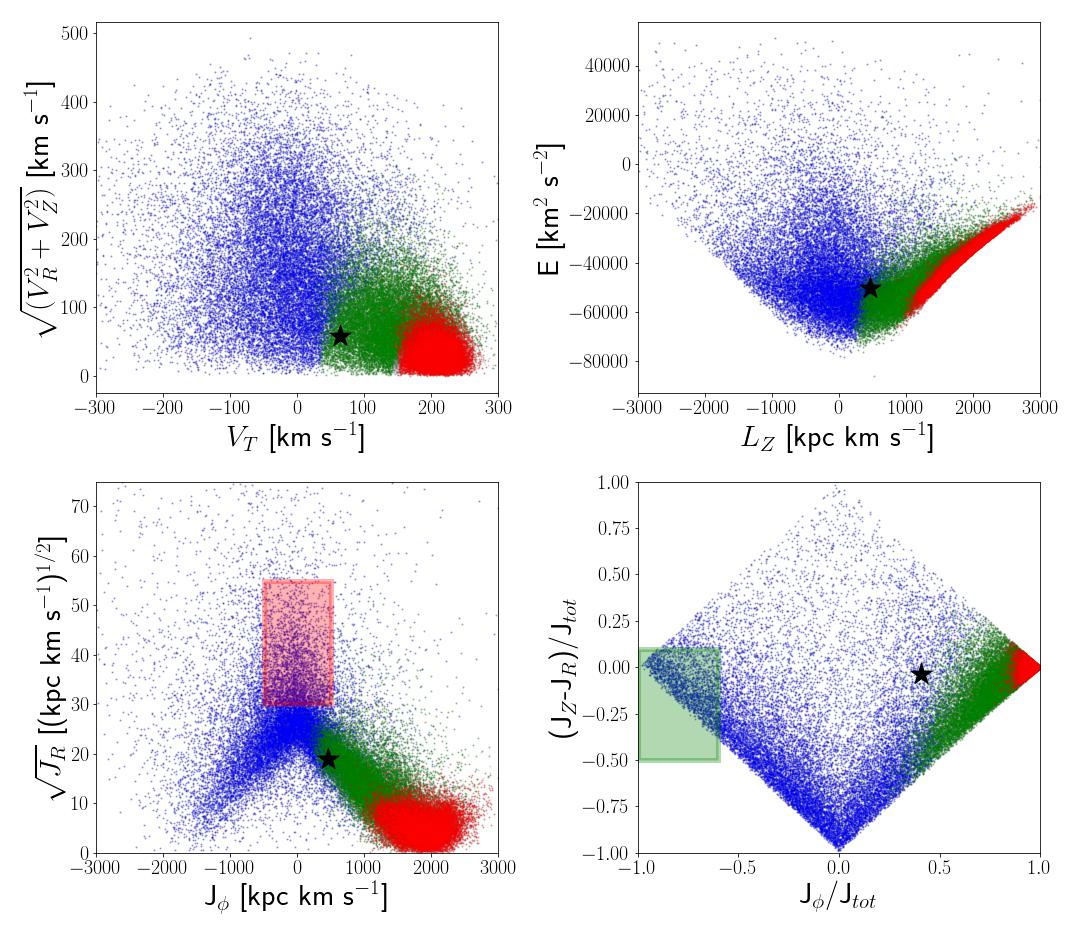}
\caption{
Upper panels: \ncstar\, position (black filled star) in the Toomre diagram (left) and the orbital energy versus angular momentum (right). Bottom panels: square root of the radial action versus the azimuthal action (left) and action diamond (right).  Coloured points are stars from the good-parallax sample of \citet[][]{bonifacio21}. The red and green shaded areas are the regions defined by \citet[][]{feuillet21} to select candidate GSE and Sequoia stars.}
\label{fig:kine}
\end{figure}

The kinematical properties of the star are shown in Fig.\,\ref{fig:kine}. The upper and lower panels display its position in several diagnostic diagrams: the Toomre diagram (top-left), orbital energy versus angular momentum (top-right), the square root of the radial action versus the azimuthal action (bottom-left), and the action diamond (bottom-right). The star is marked with a filled black star symbol. For comparison, stars from the good-parallax sample of \citet[][coloured points]{bonifacio21} are also plotted. Based on the classification criteria of \citet[][]{bensby14}, thin disc, thick disc, and halo stars are shown in red, green, and blue, respectively. The red and green shaded boxes in the bottom panels correspond to the regions defined by \citet[][]{feuillet21} for stars associated with the Gaia-Sausage-Enceladus (GSE, red box, bottom-left) and Sequoia (green box, bottom-right) accretion events.

\citet{sestito19} estimated a distance of $14.3 \pm 1.0$\,kpc and classified the system as having a retrograde Galactic halo orbit with an apocentric radius of $\rm r_{ap} = 15.9$\,kpc. We adopt a significantly shorter distance of 6.93\,kpc.
According to the \citet{bensby14} criteria, the system is classified as a thick disc star, consistent with its location in the kinematic diagrams of Fig.\,\ref{fig:kine}. However, its vertical distance from the Galactic plane ($z = -6.2$\,kpc), the maximum orbital height ($Z_{\rm max} \approx 6.3$\,kpc), and its apocentric distance ($r_{\rm ap} = 9.9$\,kpc) suggest otherwise. The orbit is eccentric ($e = 0.69$) and prograde, with positive angular momentum ($L_Z$) and transverse velocity ($v_T$). Moreover, the star does not appear to be associated with either the GSE or Sequoia structures.

We therefore conclude that \ncstar\ is best classified as a halo star on a prograde Galactic orbit.

\section{Orbital parameters}\label{sec:anal}
\begin{figure}
{\includegraphics[width=0.9\hsize, trim={0.25cm 0.cm 0.3cm 0.cm},clip]{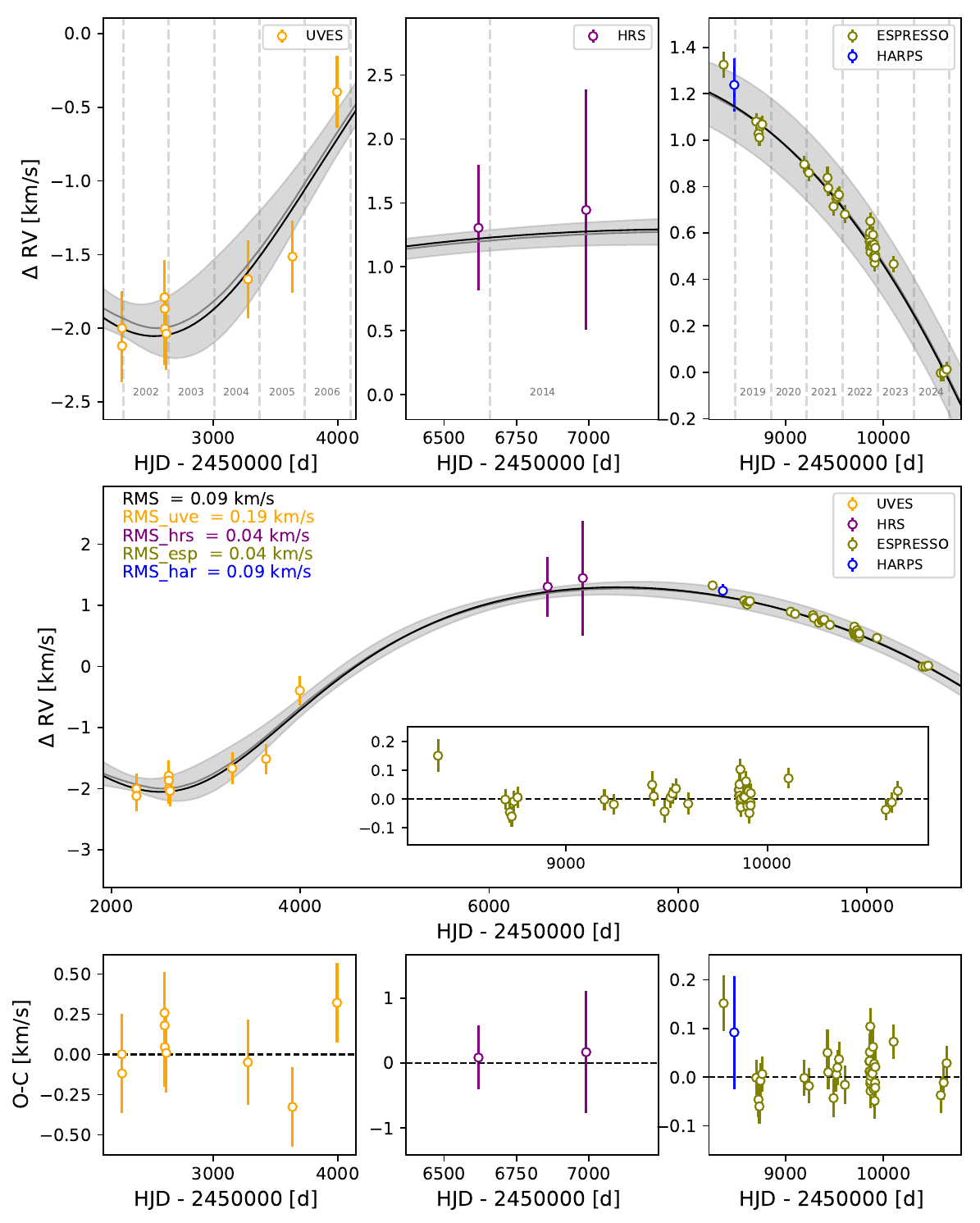}}
\caption{Radial velocities (RV)  of \ncstar\ versus heliocentric Julian date (HJD), together with the best RV model. The inner panel within the middle panel shows the ESPRESSO RV points  after removing the best  model.  The RMS of the residuals of all RVs are provided within the middle panel  and those from each spectrograph are given in the bottom panels.}
\label{vrad}
\end{figure}

We recomputed the orbital parameters of the binary star \ncstar\ considering the new measurements  
of radial velocities also for the observations of the first 3 years to benefit of a better template in the cross correlation analysis.
The new measurements from ESPRESSO spectra are provided in Table\,\ref{tab:vel}. For the UVES and HRS at SALT spectra we used
the measurements provided in \citet{bonifacio20} and \citet{arentsen19}.
For the HARPS spectrum we used that provided in \citet{aguado22}.
With respect to the epochs provided in 
\citet{aguado22}\footnote{Note that there is a typo in Table\,2 of \citet{aguado22}, the epoch 9618.061  is instead
9518.061.}, we rejected a few measurements obtained from very low S/N spectra.
The whole set of ESPRESSO radial velocities (RVs) follows a  decreasing trend  as expected  for this  long period  binary. 
The $\rm v_{rad}$ data are binned in bins of 1$-$d to minimise the stellar jitter (see Fig.\,\ref{vrad}, upper-left panel).   The RVs are modelled using  the {\sc radvel} code \citep{fulton2018radvel}\footnote{ {\sc radvel} is a Python package for modelling of radial velocity time series data, available in https://radvel.readthedocs.io/en/latest/index.html }  including a Keplerian motion of the star plus an RV offset (the $\gamma$ of the $\rm v_{rad}$ curve), and a jitter parameter for each instrument  within a likelihood scheme implemented in python using {\sc celerite}~\citep{celerite}\footnote{{\sc celerite} is a library for fast and scalable Gaussian Process (GP) Regression in one dimension, available in https://celerite.readthedocs.io/en/stable/}. 

The Keplerian orbit for radial velocity is described as:
\begin{equation}
         {\rm v_{rad}} = \gamma + k_{2} (\cos(\nu_{a} + \omega) + e \cos(\omega))
\end{equation}
where the true anomaly $\nu_{a}$ is a function of the eccentric
anomaly $E$, that is, in turn,  
a function of time, through Kepler's  equation 
\citep[see equations 1,2,3 in][]{fulton2018radvel}.
The parameters that define the orbit are five:
the orbital period ($P_{2}$), the orbital semi-amplitude velocity ($k_2$), the angle of periastron ($\omega$), the eccentricity ($e$) and the time of passage at periastron ($T_{\rm 2,peri}$). 
The {\sc radvel} program allows to infer both the time of periastron, $T_{\rm 2,peri}$, and the time at inferior conjunction of the star, $T_{\rm 2,conj}$. 
An orbit can be fit using any set of independent parameters that can be derived from these five.
Following \citet{fulton2018radvel}, we chose to perform fitting and posterior sampling using the $\ln P_2$, $T_{\rm 2,conj}$, $\sqrt{e}\cos \omega$, $\sqrt{e}\sin \omega$, $\ln k_{2}$ basis which helps to speed up the convergence. We chose $\ln k_{2}$ to avoid favouring large $k_{2}$ and $\ln P_2$ because $P_2$ may be long if compared with the observational baseline.
The $\rm v_{rad}$ errors in the Fig.\,\ref{vrad} include the jitter of each instrument and the $\rm v_{rad}$ offset has been subtracted. In addition, we adopted uniform priors for all parameters (see Table\,\ref{tab:par}).
The revised orbital period of the system is $28.9 \pm 2$\,yr. 
One should be however aware that since the time during which we have been observing this system
is shorter than the best-fit period, the actual uncertainty on the period could be larger than our estimate.

 To sample the posterior distributions and obtain the Bayesian evidence of the model (that is, marginal likelihood, ln Z), we relied on nested sampling using {\sc dynesty} \citep{speagle20dynesty}\footnote{ {\sc dynesty} is a Pure Python, MIT-licensed Dynamic Nested Sampling package for estimating Bayesian posteriors and evidences, available in https://dynesty.readthedocs.io/en/stable/ }. We initialised a number of live points equal to $N^3(N + 1)$, to efficiently sample the parameter space, with $N = 10$ being the number of free parameters.
In the middle panel of Fig.\,\ref{vrad}, we also show in grey a subsample of 300 models randomly selected from the final selection of $\sim$37,696 Bayesian samples from the posterior distributions.
The binary mass function can be computed from the masses of the binary component \citep[e.g.][]{TvdH06}, i.e. that of the seen star, $M_2$, and that of the unseen binary companion, $M_1$, and the orbital parameters as follows:
\begin{equation}
         f({\rm M}) = (M^{3}_{1} \sin^{3} i )/ (M_{1} + M_{2})^2
\end{equation}
\begin{equation}
        f({\rm M}) = [(P_{\rm orb} k_{2}^{3} )/ (2\pi G)] (1-e^{2})^{3/2}
\end{equation}

With the  Bayesian samples we computed the posterior distribution of the
binary mass function providing the following result:
\begin{equation}
   f({\rm M}) = 47_{-6}^{+8}\times 10^{-4} \, [M_{\odot}] 
\end{equation}

The binary
mass function provides a lower limit to the mass of the unseen companion.  The minimum mass of
the companion can be approximately estimated as the maximum value between $f$ and $f^{1/3} M_2^{2/3}$. Adopting a mass of $M_2[M_{\odot}]=0.8$ for the seen star we obtain $M_1 [M_{\odot}] > 0.14$   for the unseen companion. 

The minimum orbital distance (or semi-major axis) of the star with respect to the centre of mass of the binary system can be computed as:

\begin{equation}
        a_{2}\sin{i} = [(P_{\rm orb} k_{2})/(2\pi)] (1-e^{2})^{1/2} = 1.571 _{ - 0.100}^{ + 0.153}\,AU
\end{equation}

\begin{table}
\centering
	\caption{Revised orbital parameters of \ncstar\ from the Markov Chain Monte Carlo (MCMC) analysis.}
	\label{tab:par}
	\begin{tabular}{lcr}
		\hline
		Parameter & MCMC Priors & Results \\
		\hline
				&  \multicolumn{2}{c}{Keplerian orbit} \\
		\hline
		$k_2$ [km\,s$^{-1}$]  & $\mathcal{LU}$ (0.1, 5)       & 1.67$^{+0.08}_{-0.07}$    \\ 
		$P_2$ [d]           & $\mathcal{LU}$ (1000, 75000)  & 10560$^{+786}_{-518}$    \\
		$T_2-2450000$ [d]   & $\mathcal{U}$ (-4000, 3500)   &494$^{+492}_{-706}$    \\
		$\rm \omega$ [rad]  & $\mathcal{U}$ (-$\pi$, $\pi$) &  -2.87$^{+0.20}_{-0.16}$    \\
		$e$                 & $\mathcal{U}$ (0, 1)          & 0.24$^{+0.08}_{-0.08}$ \\
		\hline
		& \multicolumn{2}{c}{Other terms} \\
		\hline
		jitter$_{\rm UVES}$     [km\,s$^{-1}$] & $\mathcal{LU}$ (0.01, 5.0) & 0.24$^{+0.08}_{-0.06}$ \\
		jitter$_{\rm HRS}$      [km\,s$^{-1}$] & $\mathcal{LU}$ (0.01, 5.0) & 0.03$^{+0.08}_{-0.02}$ \\
		jitter$_{\rm ESPRESSO}$ [km\,s$^{-1}$] & $\mathcal{LU}$ (0.01, 5.0) & 0.03$^{+0.01}_{-0.01}$ \\
		jitter$_{\rm HARPS}$    [km\,s$^{-1}$] & $\mathcal{LU}$ (0.01, 5.0) & 0.05$^{+0.12}_{-0.03}$ \\
		$\gamma-46.5$           [km\,s$^{-1}$] & $\mathcal{U}$ (-3.0, 3.0)  & 0.35$^{+0.14}_{-0.17}$ \\
		\hline
	\end{tabular}
Note: the uncertainties of each quantity show the 16th and 84th percentile that covers the 68\% ($1-\sigma$) of all samples.
	\begin{minipage}{\columnwidth} 
\end{minipage}	
\end{table}

The dispersion of the ESPRESSO radial velocities after subtracting the best fit is now 47\,m\,s$^{-1}$  which is in line with the derived velocity jitter (Table~\ref{tab:par}). However, this is about 3 times larger than the mean error bar of the ESPRESSO radial velocity measurements, suggesting the presence of additional signal of relatively smaller amplitude. Pulsation, activity or the presence of a planetary motion are discussed in \citet{aguado22}.
The residuals of ESPRESSO RVs as shown in the inner panel within the centre panel of Fig.\,\ref{vrad} after subtracting the orbital solution (black line) have been analysed for periodicity and did not show any obvious peak in the periodogram.

\section{Chemical Analysis}

\subsection{Stellar parameters}\label{sec:param}
\begin{figure}
\centering
\includegraphics[clip=true,angle=0,width=0.9\hsize]{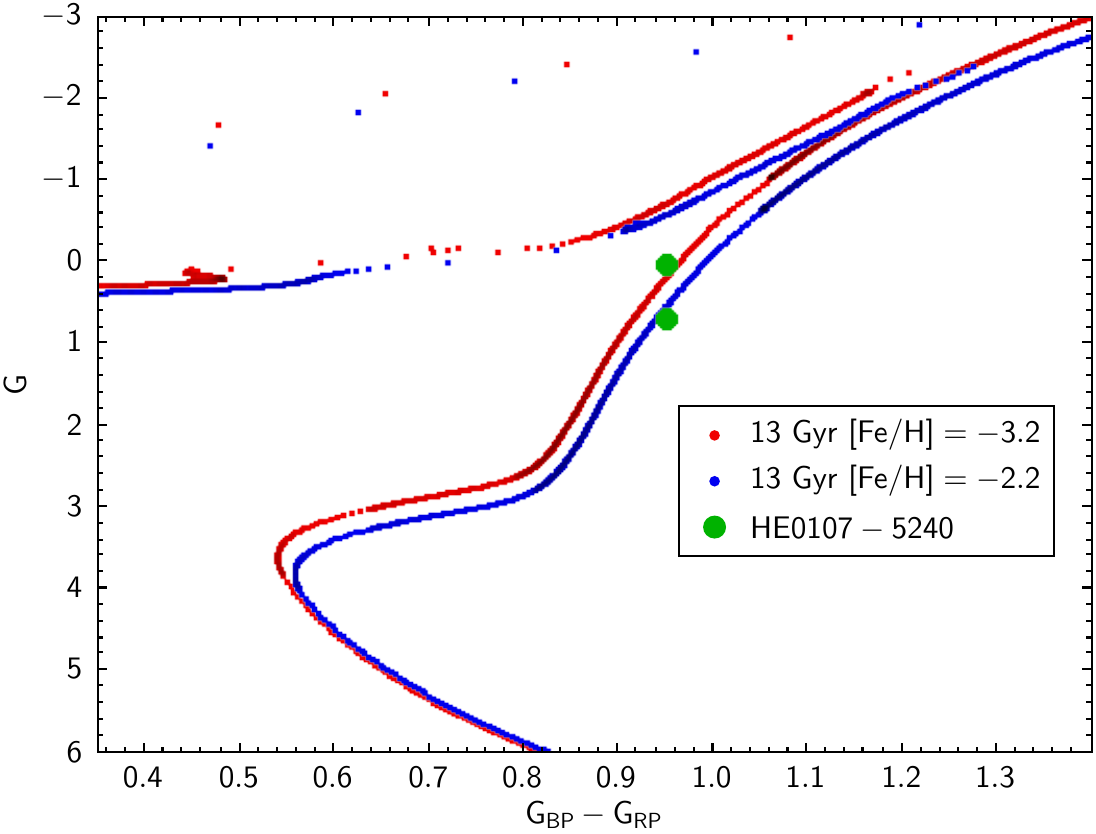}
\caption{\label{fig:isochrones} BASTI isochrones \citep{Pietrinferni_2021} in a Gaia colour-magnitude diagram. The two green points correspond to \ncstar{}, assuming surface gravities of $\log g = 2.20$ (upper point) and $2.46$ (lower point), respectively.}
\end{figure}

By using Gaia\,DR3 \citep{gaiadr3} photometry and parallax corrected by zero-point ($-0.03783$) as suggested by \citet{lindegren21}, with reddening $\rm Av=0.0344$ from \citet{schlafly11}, we derive: \teff = 5111\,K and \logg = 2.46\,[cgs], as described in \citet{leo2024}. If the zero-point correction is not applied, the derived stellar parameters would be \teff = 5116\,K and \logg = 2.20\,[cgs]. Using the former value of the surface gravity (zero-point applied), the position of the star in a Gaia colour-magnitude diagram (CMD) agrees well with a BASTI isochrone \citep{pietrinferni2021} for an age of 13\,Gyr and $\mathrm{[Fe/H]}=-2.2$ (see Fig.~\ref{fig:isochrones}). The mass range implied by the two aforementioned surface gravities is 0.77--0.78\,M$_\odot$.
In the case the zero-point is not applied, the position of the star is still compatible with the $\rm [Fe/H]=-2.2$ isochrone, but agrees better with the one of the same age and at $\rm [Fe/H]=-3.2$.
The position of \ncstar, with the parameters corrected by the zero-point, in the Gaia CMD matches only marginally with a BASTI isochrone for an age of 13\,Gyr and $\mathrm{[Fe/H]}=-3.2$ (this is the lowest [Fe/H] value provided by BASTI isochrones). 
It is true that the metallicity of \ncstar{} is $\log Z/Z_\odot \approx -2.4$, due to its very high over-abundance of CNO with respect to Fe.
The high carbon abundance leads to strong CH lines in the blue/UV spectral range, making the star to appear much redder in the $G_{\mathrm{BP}}-G_{\mathrm{RP}}$ colour compared to stars with a scaled Solar abundance pattern.
This could be an explanation why a star with $\rm [Fe/H]=-5.56$ \citep{aguado22} is compatible with an isochrone 2-3\,dex more metal-rich.
But we have to bear in mind that to build the isochrones, whatever the position in the HR diagram, all the models are computed with the same mixing length value. This is especially problematic for evolved stars \citep[see][]{manchon2024} and the fact is clear and discussed in the sample of giant stars investigated by \citet{rvs3}.

Regarding the reddening, we note that the star has a very low extinction. This is confirmed by the absence of diffuse interstellar bands (DIBs) in the high-quality ESPRESSO spectrum.
The interstellar \ion{Na}{i} doublet lines are visible but weak and their equivalent widths confirm the  low extinction. In fact the position in the sky of \ncstar\ ($\rm l=296.5^\circ$ and $\rm b=-64.49^\circ$) coincides with a well known window in the local interstellar matter where the photons from the star intercept  relatively   small clouds in the Solar vicinity.  

The stellar parameters used in the analyses of this star in the literature (see Table\,\ref{tab:abbo}) are:
\teff = 5100\,K and \logg = 2.2 \citep{christlieb02,christlieb04,bessel04,yong13,aguado22,molaro23};
\teff = 5130\,K and \logg = 2.2 \citep{collet06};
\teff = 5050\,K and \logg = 2.3 \citep{ezzeddine17};
\teff = 5300\,K and \logg = 2.5 \citep{sitnova19}.
We here adopted \teff = 5100\,K, \logg = 2.2 and  a microturbulence of 2.2\,\kms, to be consistent with the previous works, especially \citet{christlieb04}, \citet{aguado22} and \citet{molaro23}.
The two latter papers analysed a combined  spectrum derived  from the same ESPRESSO observations, but obtained  with fewer exposures.

Using the Gaia parallax, the $G$ magnitude and the bolometric correction derived
from the grid of ATLAS\,9 fluxes (Mucciarelli et al. in prep.), we estimated the luminosity of the star to  be $\rm \log (L/L_\odot) = 1.67 \pm 0.02$, where the error takes into account both the error on the parallax and on the magnitude $G$, but it is dominated by the error on the parallax.
Although red giant stars are not in a very favourable position in the Hertzsprung-Russel diagram to determine their age, we attempted to do so using the code SPInS \citep{LebretonReese2020,spins2020}.
We used a  grid of BASTI $\alpha$-enhanced  
evolutionary tracks \citep{basti}.
The input parameters were $\log (\rm T_{eff} )$, $\log (L/L_\odot)$ and $\log (Z)$.
This choice ensures that the parameters of \ncstar\ are well within
the range covered by the grid. Had we chosen [Fe/H] as metallicity indicator
the value for \ncstar\ would lie outside the grid. 
In the inference we assumed a uniform prior on age requiring that it
is smaller than the age of the Universe, 13.8 Gyr \citep{planck}.
The resulting age is 7.6\,Gyr$\pm 3.2$\,Gyr.

\subsection{Spectral energy distribution}

The VOSA service\footnote{\url{http://svo2.cab.inta-csic.es/theory/vosa}}
provides as the best fit parameters: \teff = 5500\,K and \logg = 1.5.
They indicate a considerably warmer \teff\ than our adopted one and a lower surface gravity (see Sec.\,\ref{sec:param}). 
The spectral energy distribution of the star, when compared to the flux derived from the model, does not show any UV nor IR excess.

\subsection{Abundances}\label{secabbo}

\ncstar\ is ultra-poor in Fe, but its C content is about 16\%\ solar, making this star a CEMP star. In Table\,\ref{tab:abbo} the abundances derived in the literature are shown.
As expected several upper-limits are present, but, due to the low metallicity of the star, many of them are too loose to be significant.
In particular,  the upper limits  of the heavy elements do not allow  to clarify if  \ncstar\ matches the criterion for a CEMP-no star.
Abundances or  upper-limits derived from  the new high-quality spectrum 
are provided in Table\,\ref{tab:ourabbo} and a discussion element by element is given below.
For the main 1D LTE analysis we computed spectral synthesis by using SYNTHE with an ATLAS\,9 model atmosphere \citep{K05}. In addition, we computed dedicated 3D model atmospheres with the CO5BOLD code \citep[][plus updates]{freytag2012} to estimate 3D (LTE) corrections for selected spectral lines. Details about the models and spectrum synthesis codes are provided in the appendix (\ref{sec:models}).

\paragraph{Lithium.}
No lithium from the \ion{Li}{i} doublet at 670.7\,nm
is visible. 
We applied the Cayrel's formula\footnote{$\rm \sigma_{\rm EW} = 1.6\times\sqrt{\rm PixelSize \times FWHM}/(S/N)$} multiplied by a factor three because we consider the uncertainty at $3\sigma$ \citep[see][S/N=230]{cayrel88} and we obtained $\rm A(Li)<-0.14$\,dex.

\begin{figure}
\centering
\includegraphics[clip=true,angle=0,width=0.9\hsize]{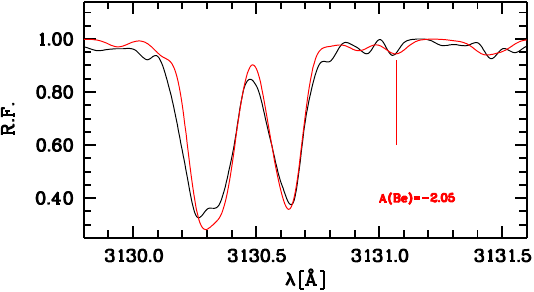}
\caption{Observed residual flux of the spectrum (solid black) in the wavelength range of the 313.1\,nm \ion{Be}{ii} resonance doublet, compared to synthesis (solid red). The position of the \ion{Be}{ii}  313.1\,nm line is marked with a vertical red line.}
\label{fig:be}
\end{figure}

\begin{figure}
\centering
\includegraphics[clip=true,width=0.9\hsize,angle=0]{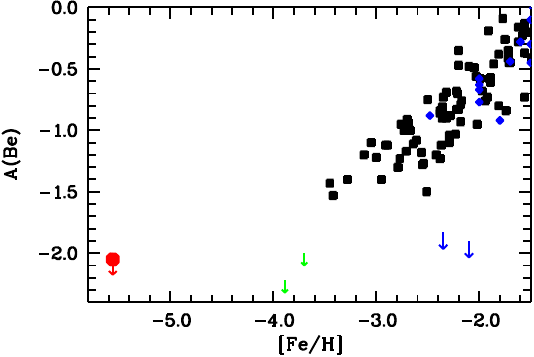}
\caption{Beryllium abundance in \ncstar\ (red symbol) compared to the Galactic behaviour Data from \citet[][black symbols]{boesgaard2011} and \citet[][blue symbols]{smiljanic2009} and the upper limits for 2MASS\,J18082002--5104378 and BD+44\,493 \citep[][green symbols]{spite2019}.}
\label{fig:fehabe}
\end{figure}

\paragraph{Beryllium.}
The two \ion{Be}{ii} resonance doublet lines lie at 313.0442 and 313.1067\,nm, both outside the wavelength coverage of ESPRESSO. The reddest \ion{Be}{ii} feature (313.1067\,nm) is not clearly detected in the UVES spectrum (see Fig.\,\ref{fig:be}), while the other line (313.0442\,nm) is heavily blended with CH molecular features. If present, the 313.1\,nm line is buried in the noise, with $\rm S/N\sim 35$. We therefore derive an upper limit (or tentative detection) of $\rm A(Be) \lessapprox -2.06$, corresponding to $\rm [Be/H] \lessapprox -3.44$ and $\rm [Be/Fe] \lessapprox +2.12$. 
However, we note that a hint of the line is present at the expected position. If confirmed in the future, this feature could potentially originate from primary spallation processes involving energetic CNO particles from the progenitor of \ncstar\ colliding with the ambient medium.
Even an upper limit slightly above $\rm [Be/Fe] < +2.0$\,dex is significant (assuming the star has not depleted its beryllium) as it argues against the existence of a primordial Be plateau (see Figure \ref{fig:fehabe}) predicted by some inhomogeneous cosmological models, albeit at lower abundance levels \citep{malaney1989,thomas1994,jedamzik2001,nakamura2017}.

\paragraph{Carbon.}
The new spectrum does not alter the derived carbon abundance, as the CH G-band features are so strong that do not benefit from an increased S/N. To determine the carbon isotopic ratio, we selected regions in the observed spectrum where clean pairs of $\rm ^{12}CH$ and $\rm ^{13}CH$ lines are present. These lines arise from the same electronic transitions, but the $\rm ^{12}CH$ features are significantly stronger due to the stellar atmosphere containing approximately two orders of magnitude more $\rm ^{12}C$ than $\rm ^{13}C$.

Our approach differs from that of \citet{molaro23} in that we used different spectral regions and simultaneously fit both the total carbon abundance and the isotopic ratio. Several wavelength intervals containing matched $\rm ^{12}CH$ and $\rm ^{13}CH$ lines were selected (5 ranges), and we performed a global fit using both parameters as free variables. This analysis yielded:
(i) a carbon abundance of $\rm A(C) = 6.72 \pm 0.02$, where the uncertainty represents the region-to-region scatter (thus, a formal error), and
(ii) an isotopic ratio of $\rm ^{12}C/^{13}C = 110 \pm 12$.
Our carbon abundance is in excellent agreement with the value of $\rm A(C) = 6.75$ reported by \citet{molaro23}, and the isotopic ratio is consistent within 1.7$\sigma$ with their result of $\rm ^{12}C/^{13}C = 87 \pm 6$.

To evaluate the impact of 3D effects, we computed the 3D (LTE) correction for the $\rm ^{12}CH$ line at 423.1\,nm, obtaining a correction of $-0.65$\,dex for the $\rm ^{12}C$ abundance. To assess the 3D correction of the isotopic ratio, we focused on the same spectral region, examining pairs of $\rm ^{12}CH$ and $\rm ^{13}CH$ lines sharing the same electronic transition. While both lines experience similar 3D effects at equal line strengths, in practice their strengths differ significantly. For a solar carbon isotopic ratio, the equivalent width (EW) of the $\rm ^{12}CH$ line is tens of times larger than that of the corresponding $\rm ^{13}CH$ line.

\begin{figure}
\centering
\includegraphics[width=0.9\hsize,clip=true]{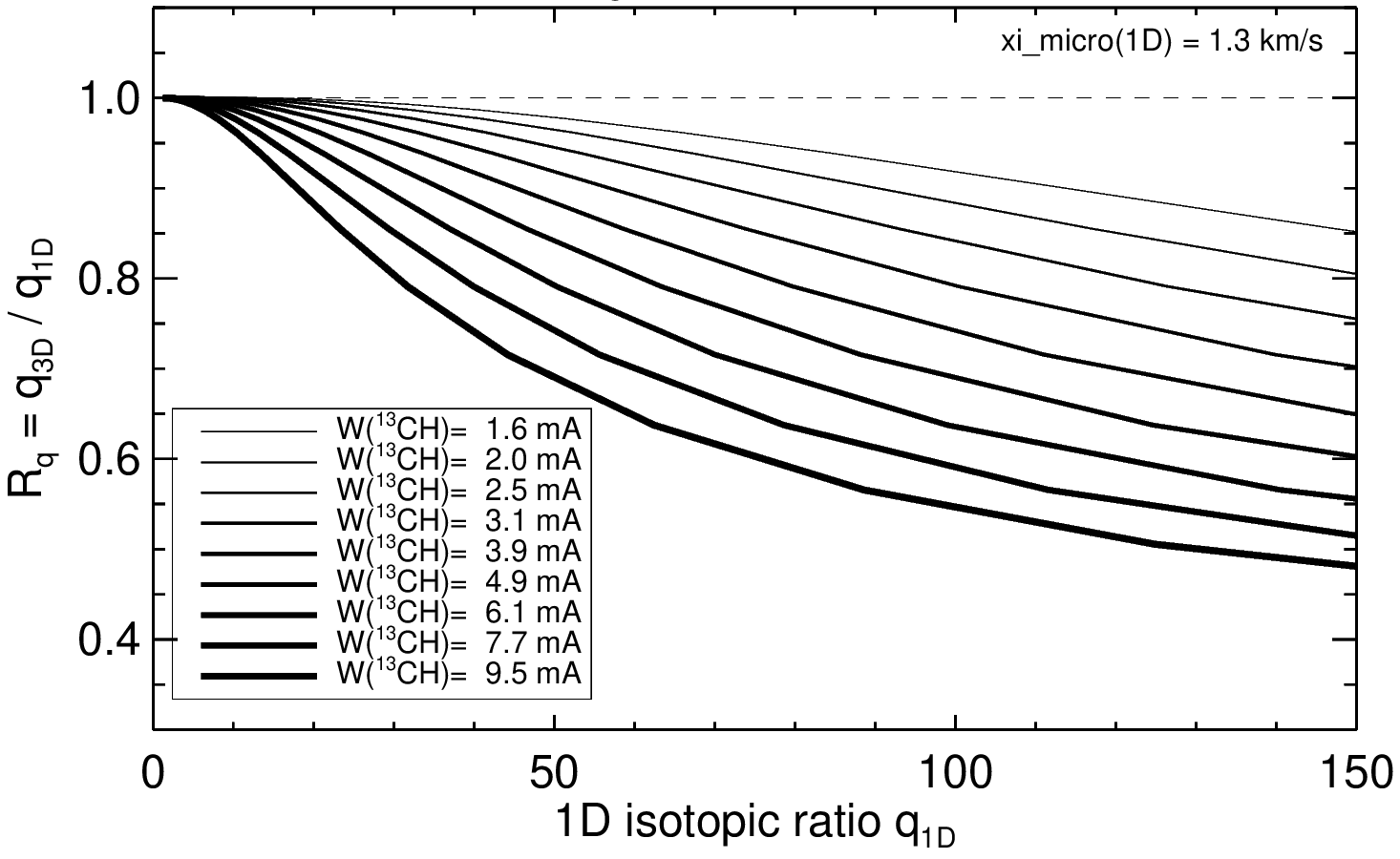}
\caption{3D correction factor for the $^{12}$C/$^{13}$C isotopic ratio versus the 1D isotopic ratio for different strengths of the $\rm  ^{13}CH$ component of the 423.1\,nm line pair. According to this diagram, a 1D isotopic ratio of 100 with a $^{13}$CH equivalent width of 2\,m\AA\ would translate to a 3D isotopic ratio of $\approx 90$.}
\label{fig:3dqciso}
\end{figure}

We computed the 3D correction factor of the $\rm ^{12}C/^{13}C$ isotopic ratio, $R_{\rm q}$, as the ratio of the isotopic ratios derived in 3D and in 1D, $R_{\rm q} = q_{\rm 3D}/q_{\rm 1D}$ and expressed this as a function of $q_{\rm 1D}$ for different choices of the $\rm ^{13}CH$ line’s equivalent width, $\rm EW_{13}$, as shown in Fig.\,\ref{fig:3dqciso}. The correction is also sensitive to the adopted microturbulence. We used a value of 1.3\,\kms, selected to ensure a flat 3D abundance correction with the strength of the 516.9\,nm \ion{Fe}{ii} line (see below).

Unfortunately, the derived 3D corrections depend on the chemical composition,
in particular the C, N, O abundances, adopted for the construction of the 3D
atmosphere and the 1D reference model. All 3D results reported above are based
on a 3D (and 1D reference) model with a solar chemical composition scaled down
by $-4.0$\,dex, except for the $\alpha$ elements (including O but not C) that
are enhanced by $+0.4$\,dex relative to the other metals. Compared to the
abundances derived for HE\,0107--5240 in this work, C and O are too low by
about $-2.3$ and $-0.65$\,dex, respectively, in this model.

As an alternative, we also computed a pair of 3D -- 1D models where the carbon and nitrogen abundances are enhanced by $+3.0$\,dex and the oxygen abundance by $+2.6$\,dex with respect to the previous case. Again, these C, N, O abundances are not representative of HE\,0107--5240, now C and O are too high by about $0.7$ and $2.0$\,dex, respectively. The C/O ratio is $\sim 0.5$ instead of $\sim 10$.

With the same procedure, we derived the 3D correction for the C abundance and the 3D correction for the $^{12}$C/$^{13}$C isotopic ratio. As already found by \citet{gallagher2017}, the 3D correction is smaller (only about $-0.2$\,dex) and the 3D correction factor for the C isotopic ratio increases from $\sim 0.8$ to $\sim 1.1$, implying that the 3D corrected isotopic ratio is larger than
what is derived in 1D.  What is important to underline here is that a robust result, beyond all uncertainties of the modelling, is that HE\,0107--5240 has a high $^{12}$C/$^{13}$C isotopic ratio ($\ga 90$), ruling out any mixing with CNO processed material.

More reliable 3D corrections can only be obtained from a future 3D model based on custom-made opacity tables with consistent CNO abundances.

\paragraph{Nitrogen.}
The NH band at 336\,nm is outside the wavelength coverage of ESPRESSO.
We then used the UVES 346 spectrum to investigate it. We inspected this 336\,nm NH band by using the molecular data provided by \citet{fernando2018} and we derived A(N)=4.42  
($\rm [N/H]=-3.44$ and $\rm [N/Fe]=+2.12$).
The 3D correction on the N abundance from this band is $-0.58$\,dex, that, if applied, would decrease the [N/Fe] ratio to 1.6\,dex.

\paragraph{Oxygen.}
No atomic or molecular useful line is available in the wavelength range of ESPRESSO.
The OH lines in the UV range of a UVES spectrum have been already investigated by \citet{bessel04}. 
We selected the OH lines in the list suggested by \citet{prakapavicius2017}, as the molecular data we adopted are from their table\,A.1. We investigated ten OH lines and we derived $\rm\langle A(O)\rangle =5.72\pm 0.09$.
With a solar oxygen abundance of 8.76 (see Table\,\ref{tab:ourabbo}), we obtained $\rm [O/H]=-3.04$ and $\rm [O/Fe]=+2.52$\,dex.
The 3D effects on molecular lines are strong. The average corrections we derive is $-0.58\pm 0.07$\,dex,
but we have to stress that the corrections have been computed without taking into account scattering in the synthesis of both 1D and 3D profiles.

\paragraph{Sodium.}
The two \ion{Na}{i}-D resonance lines are strong in the spectrum, providing an abundance of A(Na)=1.64.
According to \citet{andrievsky2007}, NLTE corrections for the \ion{Na}{i} D lines are small at extremely metal-poor regime.
Two  weak features due to Galactic interstellar absorption, account for the very low reddening. 

\begin{figure}
\centering
\includegraphics[width=0.9\hsize,clip=true]{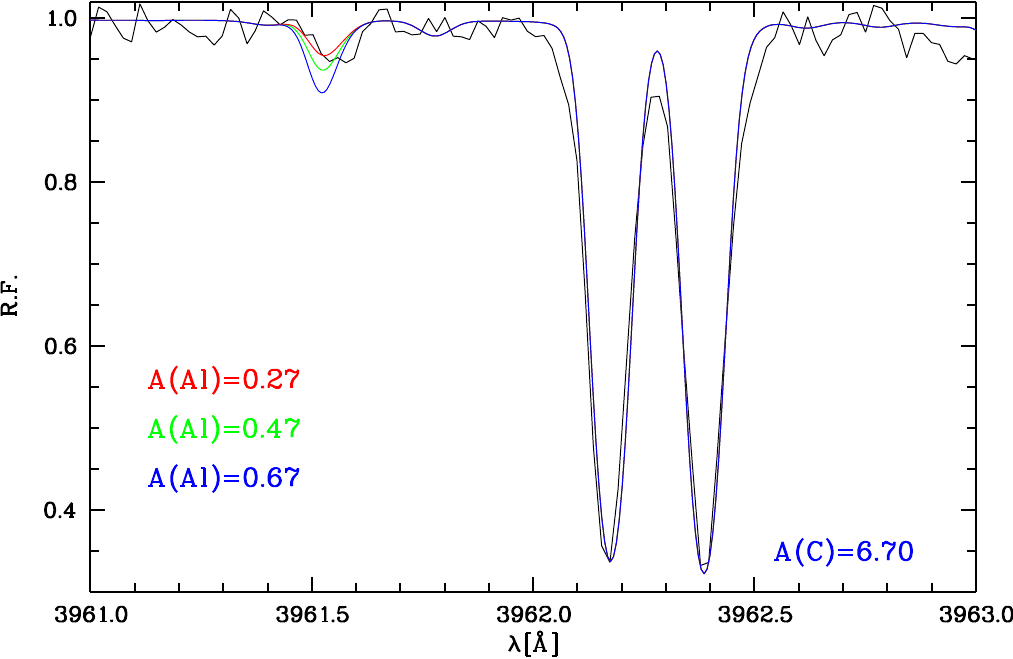}
\caption{Observed spectra (solid black) in the range of the \ion{Al}{i} 396.1\,nm  line:
 tentative detection of Al (solid red) and other syntheses (solid blue) to show the Al line.}
\label{fig:al}
\end{figure}

\paragraph{Aluminium.}
At the wavelength of the strongest \ion{Al}{i} line at 396.1\,nm, a feature compatible with the Al line is visible. However,
on the red side there is a spike due to noise that makes the detection uncertain.
Either this is a tentative Al detection with an abundance of $\rm A(Al)=0.27$;\ or we have a conservative  higher upper limit at $\rm A(Al)<0.47$. In both cases  the star is poor in Al with $\rm [Al/Fe]=-0.64$:\ or $\rm [Al/Fe]<-0.44$ (see Fig.\,\ref{fig:al}).
The Al resonance lines are affected by strong NLTE effects. For the well known star CD--38:245, which  has  stellar parameters similar to \ncstar,  \citet{andrievsky2008} derived a NLTE correction of $+0.74$\,dex. 
The NLTE correction for Fe, according to the MPIA website\footnote{\url{https://nlte.mpia.de/gui-siuAC_secE.php}} (hereafter MPIA), is about +0.55\,dex (see below the paragraph on iron).
So, taking into account the NLTE corrections for Al and Fe, the star is compatible to be solar in the [Al/Fe] ratio.

\paragraph{Silicon.}
The strong \ion{Si}{i} line at 390.5\,nm is not clearly visible in the spectrum, but there is a hint of a line  that could be attributed to the  \ion{Si}{i} line. We  provide a conservative upper limit of $\rm A(Si)< 1.8$ 
corresponding to $\rm [Si/Fe]< -0.16$. 
In fact, we guess a feature, compatible with the noise, at the expected wavelength, corresponding to a lower Si abundance ($\rm A(Si)\sim 1.6$, $\rm [Si/Fe]\sim -0.36$). 
For this line, the NLTE corrections provided in the web-site MPIA 
are large (+0.31\,dex) when adopting a metallicity of $-5.0$ for the star, but still smaller than the NLTE correction of Fe. At metallicity $-4.0$ the NLTE correction for both elements are much smaller (+0.1\,dex for Si and +0.3\,dex for Fe).

\paragraph{Sulphur.}
The strongest \ion{S}{i} feature available in the ESPRESSO spectrum is the strongest triplet of the Mult.\,8 at 675.7\,nm. However, this feature is already weak in solar-metallicity or slightly metal-poor stars and
the upper-limit we derive ($\rm A(S)<5.2$) 
is not significant.
The  UVES spectra with setting 860\,nm  host the stronger \ion{S}{i} lines of Mult.\,1. We selected  the \ion{S}{i} lines at 922.8 and 923.7\,nm, which are free from telluric absorption. The averaged spectrum from 4 OBs provided a much more stringent upper-limit at $\rm A(S)<4.40$ ($\rm [S/Fe]<2.80$) but still non-significant. 
The \ion{S}{i} line of Mult,\,1 are very sensitive to NLTE effects. According to \citet{takeda2005}, for the line at 923.7\,nm a NLTE correction of about $-0.5$\,dex is expected for a metallicity of $-4.0$.

\paragraph{Potassium.}
The strong \ion{K}{i} line at 769.8\,nm is not visible in the spectrum of \ncstar\ and no telluric contamination affect the position of the line.
We took into account the S/N ratio of 200 and using Cayrel's formula we derived an upper-limit of $\rm [K/Fe]<1.50$, 
which is not significant.
Looking at the computations from \citet{andrievsky2010}, the NLTE correction for this line is not large, of the order of +0.1\,dex.

\paragraph{Scandium.}
Two \ion{Sc}{ii} lines (361.38 and 363.07\,nm) are available in the UVES 346 spectrum. In spite of the low  S/N ratio in this spectral region 
we have been able  to detect the 361.3\,nm \ion{Sc}{ii} line and we derived $\rm A(Sc)=-2.3$ ($\rm [Sc/H]=-5.40$).

\paragraph{Chromium.}
Due to the good spectral quality (S/N of almost 100), at 425.43\,nm  a feature is visible in the spectrum.
According to the synthesis this is a blend of the \ion{Cr}{i} line at 425.43\,nm with two CH lines, so the contribution is due to CH. 
In the green wavelength range, at 520\,nm, the S/N is almost 200 and three \ion{Cr}{i} lines are not visible. By using Cayrel's formula, we derived an upper-limit too small, that we think too optimistic for such weak lines. 
We think that an upper-limit of $\rm A(Cr)<0.30$ and $\rm [Cr/Fe]<0.22$ is reasonable.
The Cr upper-limit is not significant to draw conclusions on the star.
On top of that, the NLTE correction is large and positive (of +0.8\,dex in the case of 
metallicity $-4.0$) according to MPIA. 
This value is larger than the NLTE correction for Fe, fact that makes this upper-limit non significant.

\paragraph{Manganese.}
The \ion{Mn}{i} at 403.3\,nm is not visible, providing (using Cayrel's formula) a stringent upper-limit ($\rm A(Mn)<-0.50$.
But the NLTE correction expected from these line is large \citep[$\sim 1$\,dex, see][]{bergemann08mn}.

\begin{figure}
\centering
\includegraphics[width=0.9\hsize,clip=true]{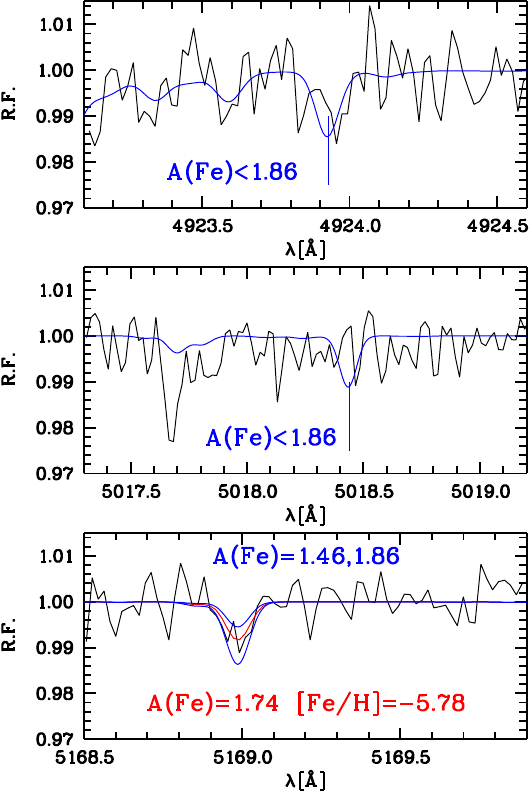}
\caption{Observed spectra (solid black, from top to bottom  \ion{Fe}{ii} 492.3, 501.8 and 516.9\,nm), compared to the best fit (solid red) and synthetic profile to provide uncertainty and upper-limit (solid blue).}
\label{fig:fe2}
\end{figure}

\paragraph{Iron.}
We keep A(Fe) from neutral lines from \citet[][A(Fe)=1.96]{aguado22} because the value we derived from the latest ESPRESSO combined spectrum is just 0.005\,dex higher.
\citet{christlieb04} provided upper limits on A(Fe) from two \ion{Fe}{ii} lines at 501.8440 and 516.9028\,nm \citep{nave2013}. Of the two
lines the 516.9\,nm line is weak but clearly visible in the ESPRESSO spectrum. We adopted $\rm\log{gf}=-1.0$ from \citet{melendez2009}, while \citet{schnabel2004} provide a \loggf\ of $-0.87$. We derived A(Fe)=1.74 from line-profile fitting (see Fig.\,\ref{fig:fe2}).
From the other two \ion{Fe}{ii} lines expected to be strong enough to be visible in extremely metal-poor stars (492.3921 and 501.8440\,nm), we derived just an upper-limit, consistent with the abundance from the 516.9\,nm line (see Fig.\,\ref{fig:fe2}).
As expected, according to MPIA, 
the NLTE correction for the 516.9\,nm line is very small (0.03\,dex).
Our 3D LTE model provides a 3D correction for this line of about +0.1\,dex.
In the UVES setting 346,
we detected also  the \ion{Fe}{ii} line at 322.774\,nm for which we derived A(Fe)=1.67 (with a \loggf\ of $-1.014$), in good agreement with the \ion{Fe}{ii} line in the ESPRESSO spectrum.

Selecting among the \ion{Fe}{i} lines investigated by \citet{aguado22}, 12 are present in the database MPIA
to derive 1D NLTE correction for iron.
From these lines, we derived a NLTE correction of +0.54, adopting a model with parameters: \teff=5100\,K, \logg=2.2, $\rm [Fe/H]=-5.0$ and microturbulence of 2.2\,\kms\ ($-5.0$ is the lowest metallicity provided).
By applying the NLTE correction to the A(Fe)=1.96 derived from \ion{Fe}{i} lines, we have a disagreement with the abundance from the A(Fe)=1.74 \ion{Fe}{ii} line.
The NLTE correction from the same \ion{Fe}{i} line adopting a metallicity $-3.0$, compatible with the oxygen abundance, is +0.12\,dex, and with a metallicity $-2.0$, compatible with the C content, is +0.08\,dex.
Clearly, a  `high' NLTE correction in A(Fe) from neutral lines brings tension with the Fe abundance derived from the ionised lines.

\paragraph{Cobalt.}
The \ion{Co}{i} UV lines (345.35 and 352.98\,nm) are outside the wavelength range of ESPRESSO, but they are in the UV UVES range. 
The result from the UVES spectrum is not so clear. (i)~The feature at 345.35\,nm is compatible with $\rm A(Co)\sim -0.30$. The line is blended with a non-reproduced NH line, but the blend is not affecting the abundance derived. (ii)~The two \ion{Co}{i} lines at 352.9\,nm are consistent and provide $\rm A(Co)\sim 0.30$ and $\rm [Co/H]=-4.62$.

\paragraph{Zinc.}
The line at 481.5\,nm is the strongest \ion{Zn}{i} line available in the wavelength range provided by ESPRESSO.
We did not detect the line in the spectrum because it is too weak. 
Any upper-limit is not stringent and Cayrel's formula was too optimistic, so we over-plotted synthetic spectra and we derived an upper-limit on Zn for $\rm A(Zn)<1.02$ 
which is lower than the previous values provided in the literature (see Table\,\ref{tab:abbo}) but sill not significant.
Looking in the tables by \citet{takeda2005}, the NLTE correction is positive but not large ($\sim 0.1$\,dex).

\paragraph{Strontium.}
The 407.7\,nm \ion{Sr}{ii} line is not detectable in the spectrum.
We computed the S/N ratio as 90 and, using Cayrel's formula \citep[multiplied by three,][]{cayrel88}, we derive $\rm A(Sr)<-3.89$ ($\rm [Sr/Fe]<-1.25$), which is a very stringent upper-limit.
Looking at a star (CD--38:245) with similar parameters investigated by \citet{francois2007}, \citet{andrievsky2011} provide a NLTE correction of +0.23\,dex. Also taking into account the NLTE correction, the star is Sr-poor, moreover considering the NLTE correction for Fe from neutral lines provided in the literature (and in contradiction with our results on \ion{Fe}{ii}) is positive and expected to be larger.

\begin{figure}
\centering
\includegraphics[width=0.9\hsize,clip=true]{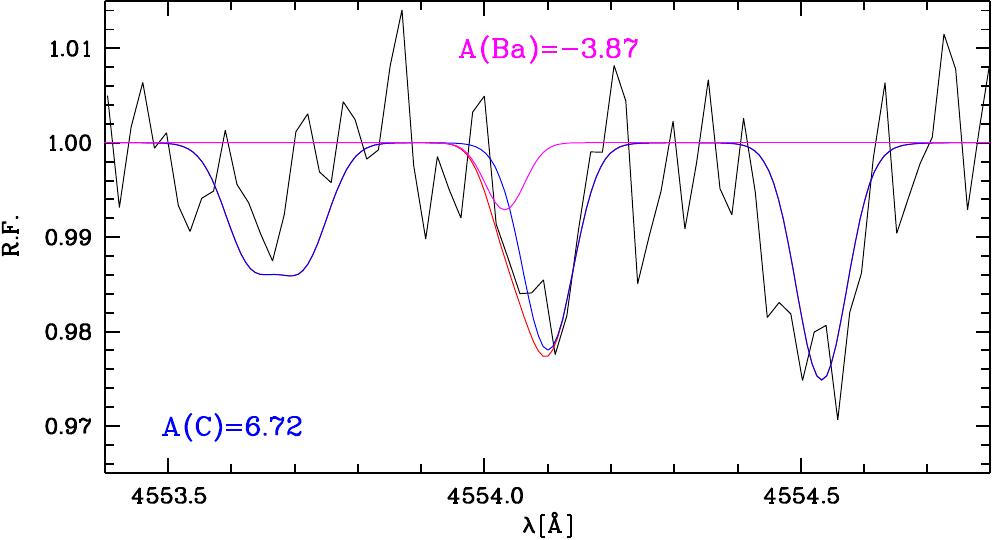}
\caption{Observed spectra (solid black) in the range of the \ion{Ba}{ii} 455.4\,nm  line compared to
 synthesis of the upper-limit on Ba (solid pink), synthesis on the contribution of CH (sold blue) and synthesis with CH and Ba upper-limit (solid red).}
\label{fig:ba}
\end{figure}

\paragraph{Barium.}
Deriving a Ba detection or a stringent upper-limit is extremely important in the case of CEMP stars  to understand to which group they belong. 
A low Ba abundance or upper-limit providing $\rm [Ba/Fe]<0$ would imply that the star is a CEMP-no.
The \ion{Ba}{ii} line at 455.4\,nm is blended in this C-rich star with CH molecules, that have to be taken into account.
The CH line shows an asymmetry that could be due to the Ba contribution. The S/N ratio per pixel in the wavelength range is 160, that, by using Cayrel's formula, provides at $3\sigma$ $\rm A(Ba)<-3.87$ ($\rm [Ba/H]=-6.04$ and $\rm [Ba/Fe]=-0.48$ when using A(Fe) from \ion{Fe}{i} lines).
The asymmetry in the CH line is more than justified by this upper-limit. In Fig.\,\ref{fig:ba} we provide the observed spectrum compared to the synthesis with the CH contribution, with the upper-limit on Ba and with both.
Considering the 1D-NLTE correction for CD--38:245 \citep{francois2007}, \citet{andrievsky2011} provide $+0.32$\,dex, but the NLTE of A(Fe) from \ion{Fe}{i} lines is expected to be larger.

\paragraph{Europium.}
The \ion{Eu}{ii} line at 412.973\,nm is not visible in the spectrum.
In the wavelength range we measured a S/N around 80 that, by using Cayrel's formula at $3\sigma$, implies an upper-limit of $\rm A(Eu)<-3.0$ but few spike of noise pushed us to adopt $\rm A(Eu)<-2.67$.
Both values are not significant.

\section{Discussion}

\ncstar\, is a Halo binary system with a prograde Galactic orbit. 
Its abundances reflect the chemical composition of the cloud from where the system was formed. 
The star belong to a binary system with an estimated period of about 29\,y.

\subsection{Stellar chemistry}
\subsubsection{The carbon isotopic signature: high $\rm ^{13}C$ abundance }
Notably, the absolute abundance of $\rm ^{13}C$ in \ncstar, $\rm A(^{13}C) = 4.68$, exceeds that of nitrogen ($\rm A(N) = 4.42$) and is second only to oxygen ($\rm A(O) = 5.72$) and $\rm ^{12}C$ ($\rm A(^{12}C) = 6.72$), underscoring the significant presence of $\rm ^{13}C$ in the stellar atmosphere.

The carbon isotopic ratio ($\rm ^{12}C/^{13}C$) measured in \ncstar\, (also accounting for 3D corrections) is  consistent with the value reported by \citet{molaro23} and slightly  higher than those observed in the sample of unmixed giant stars analysed by \citet{spite06}.

Nonetheless, the derived ratio remains well below the predictions of the zero-metallicity stellar models, which typically yield $\rm ^{12}C/^{13}C$ ratios exceeding 1000 \citep[e.g.][]{roberti+2024}. 
Ratios in the range $\rm ^{12}C/^{13}C = 50$–100 can be produced when the hydrogen- and helium-burning shells merge during the pre-supernova evolution of massive stars. This scenario is exemplified by the $\rm 25 M_\odot$ zero-metallicity model with an initial rotational velocity of 300\, km s$^{-1}$ presented by \citet{roberti+2024}. Incorporating the nucleosynthetic yields from such models into a Galactic chemical evolution framework, \citet{rizzuti2025} successfully reproduced observed abundance patterns (provided that H-He shell merging also occurs in stars with metallicities up to $\rm [Fe/H] = -3.0$).
The carbon isotopic ratio observed in \ncstar\ also rules out self-pollution from an AGB companion as the origin of its carbon enhancement, as such a process would produce ratios of $\rm ^{12}C/^{13}C < 10$. Instead, the value is consistent with those found in other unevolved carbon-enhanced metal-poor stars, such as HE\,1327$-$2326 and HE\,0233$-$0343 \citep[see][]{molaro23}.
It is important to note, however, that measuring isotopic ratios above $\sim 100$ in metal-poor stars is observationally challenging, if not unfeasible. Therefore, published lower limits below this threshold may still be compatible with much higher intrinsic values, potentially exceeding 1000.

\subsubsection{Reevaluating NLTE effects: iron lines show LTE consistency}
\label{sec:discussfe2}

The LTE iron abundances derived from both neutral and ionised lines in \ncstar\ are in perfect agreement within the uncertainties \citep[as already noted by][]{christlieb2008}, without applying any corrections, contrary to what NLTE studies typically predict. The \ion{Fe}{ii} line at 516.9\,nm forms at approximately $\log \tau \sim -1$, relatively deep within the stellar atmosphere. Since \ion{Fe}{ii} is the dominant ionisation state of iron under these conditions, this line is expected to form close to LTE. The agreement between \ion{Fe}{i} and \ion{Fe}{ii} abundances thus implies that NLTE effects on \ion{Fe}{i} must be small for \ncstar.

This finding stands in contrast with the literature, where significant positive NLTE corrections are generally reported for neutral iron lines in extremely metal-poor stars \citep[see, e.g.,][]{amarsi2016,ezzeddine17}. For example, \citet{nordlander2017} investigated 3D-NLTE effects on iron in the ultra iron-poor star SMSS\,0313--6708, which has stellar parameters nearly identical to those of \ncstar; in fact, their model was computed specifically for \ncstar. They found a 3D-NLTE correction of approximately +0.8\,dex.

If we were to force the \ion{Fe}{ii} abundance to match the lower \ion{Fe}{i} abundance derived by \citet{ezzeddine17} under NLTE assumption, we would need to increase the surface gravity of \ncstar\ by more than 2\,dex, effectively classifying it as a Turn-Off star. However, this scenario is clearly inconsistent with the Gaia parallax and with the observed profiles of H$\alpha$ and other Balmer lines, whose wings are too narrow to support such high gravity.

This unexpected result for \ncstar, which is supported by a preliminary consistent result for SMSS\,J160540.18--144323.1 \citep{nordlander2019} discussed in the Appendix\,\ref{sec:nordlander}, challenges current expectations and highlights the need for further investigation into 1D- and 3D-NLTE effects in the most iron-poor stars.

\subsubsection{Lithium depletion}

\begin{figure}
\centering
\includegraphics[width=0.9\hsize,clip=true]{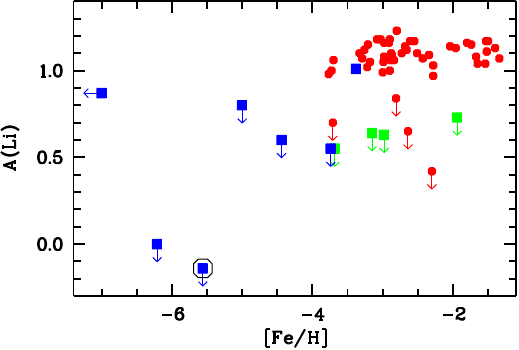}
\caption{Update figure of Figure\,2 from \citet{mucciarelli22}, with same symbols: red filled circles normal star, blue and green filled squares CEMP stars with `low' and `high' A(C), respectively. \ncstar\ is highlighted by a black circle.}
\label{fig:amplateau}
\end{figure}

According to its stellar parameters, \ncstar\ should lie on the Mucciarelli plateau \citep[][]{mucciarelli22}, which implies an expected lithium abundance of $\rm A(Li) \sim 1.09$. 
In Fig.\,\ref{fig:amplateau} we reproduce Figure\,2 from \citet{mucciarelli22}, adding SMSS\,J1505--1443 \citep{nordlander2019} and SDSS\,J1313--0019 \citep{allendeprieto2015} and updating value for \ncstar.
The upper limit on lithium ($\rm A(Li) < -0.14$) of \ncstar\ is consistent with an absence of lithium, suggesting that either the star has destroyed its lithium or it formed from a gas cloud already devoid of lithium.

It is true that several CEMP-no stars fall below the Mucciarelli plateau \citep[see Figure~2 in][]{mucciarelli22}, yet some do lie on the plateau. For example, \citet{matsuno17} measured lithium at the Spite plateau level in two CEMP-no dwarf stars with metallicities around [Fe/H] $\sim -3.0$. This indicates that the nature of their progenitors alone cannot explain the higher incidence of lithium depletion at low metallicities. As discussed in \citet{molaro23}, the mechanism responsible for lithium depletion appears to correlate more strongly with extremely low iron abundance, rather than with destruction during the evolution of faint supernovae, the likely progenitors of CEMP-no stars.

An alternative explanation is that \ncstar\ is an evolved blue straggler, formed through the merger of two lower-mass stars, during which lithium could have been destroyed. Interestingly, the relatively young age we derived for \ncstar\ — unusual for a halo star — may support this hypothesis. However, given the large uncertainty in the age estimate, it remains consistent within $1.8 \sigma$ with an age equal to that of the Universe.

\subsubsection{n-capture elements}

\begin{figure}
\centering
\includegraphics[width=0.9\hsize,clip=true]{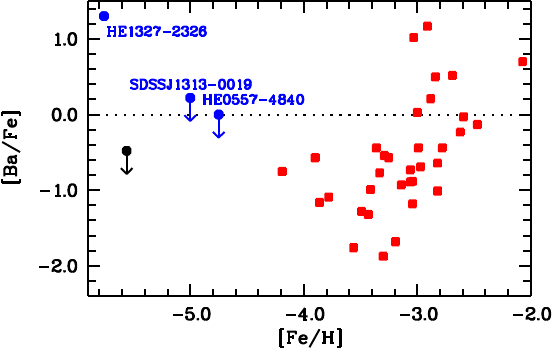}
\caption{[Ba/Fe] versus [Fe/H] (black arrow) compared to the sample in \citet{francois2007} (red full red).
Three CEMP stars HE\,1327--2326 \citep{molaro23} SDSSJ\,131326.89--001941.4 (Frebel et al. 2015, ApJ 810,27) and HE\,0557--4840 (Norris et al. 2007, ApJ 670, 774) are added as blue symbols.}
\label{fig:bafe}
\end{figure}

\begin{figure}
\centering
\includegraphics[width=0.86\hsize,clip=true]{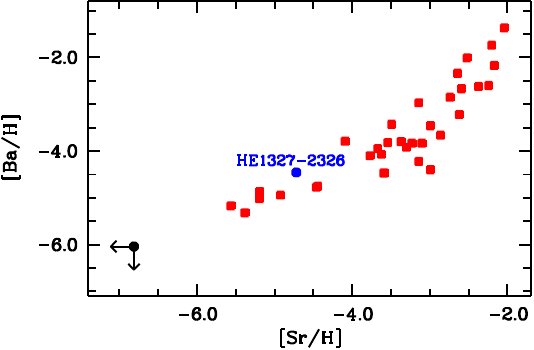}
\caption{[Ba/H] versus [Sr/H] (double black arrow) compared to the sample in \citet{francois2007} (red full red) and to HE\,1327--2326 \citep[][blue filled circle]{molaro23}.}
\label{fig:srba}
\end{figure}

The new upper limit for barium in \ncstar\ satisfies the criterion for classification as a CEMP-no star. In Fig.\,\ref{fig:bafe}, the star is compared to a sample of EMP stars (excluding CEMP stars) analysed under LTE assumption by \citet{francois2007}, and it displays a comparable [Ba/Fe] ratio. However, a striking contrast is seen with HE\,1327--2326, which shows a high barium abundance of [Ba/Fe] = +1.0 \citep{molaro23}, highlighting the large diversity in the behaviour of neutron-capture elements at low metallicities.
Also in the [Ba/H] vs. [Sr/H] plane, \ncstar\ is consistent with metal-poor stars (see Fig.\,\ref{fig:srba}).

\citet{thansen15} proposed the existence of a floor in the absolute barium abundance at $\rm A(Ba) \sim -2.0$ for CEMP-no stars. However, \ncstar, with its extremely low [Fe/H] and an upper limit of $\rm A(Ba) < -3.87$, lies well below this proposed floor and below all upper limits reported in their study (see their Figure 13). This suggests that either the barium floor occurs at a lower level than previously thought, or it may not exist at all.

\subsection{Origin of \ncstar}

\citet{limongi+2003} proposed a model to explain the observed abundance ratios in \ncstar\ by invoking the ejecta of primordial core-collapse supernovae. Specifically, they suggested that elements heavier than Mg were synthesised in one or more `normal' Population III core-collapse supernovae, while the lighter elements originated from enrichment by a failed Pop\,III supernova - i.e., one that left behind a very massive remnant. This scenario was based on a detailed analysis of presupernova models and explosive yields published by \citet{CL02} and \citet{LC02}.

In particular, they found that stars with initial masses around $\rm 35~M_\odot$ underwent partial mixing between the He convective shell and the tail of the H shell. This mixing, which is relatively common in zero-metallicity massive stars \citep[see, e.g.,][]{fujimoto1990,chieffi+2001,LC12,roberti+2024}, leads to the synthesis of C, N, Na, and Mg in relative proportions consistent with those observed in \ncstar.

A quantitative analysis using these models demonstrated that the observed abundance pattern (specifically, the elements C, N, Na, Mg, Ca, Ti, and Ni) could be reproduced by combining the yields of a $\rm 15M_\odot$ `normal' Pop\,III supernova with those of a $\rm 35M_\odot$ model in which the mass cut was located within the He shell, corresponding to a remnant mass of $\rm 9.4~M_\odot$ \citep[see Figure 3 in][]{limongi+2003}.

\begin{figure}
\includegraphics[width=7.0truecm,clip=true]{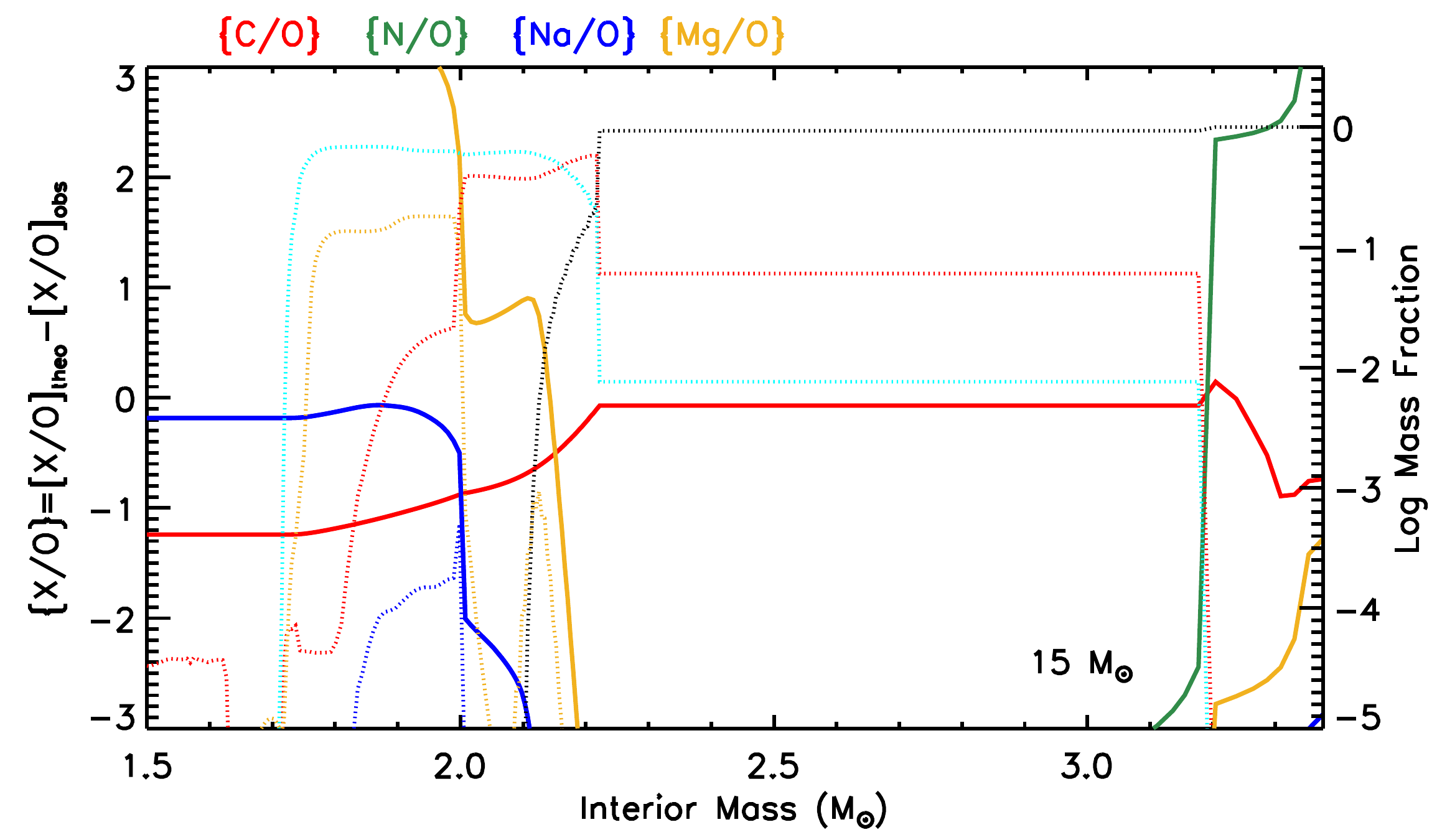}
\includegraphics[width=7.0truecm,clip=true]{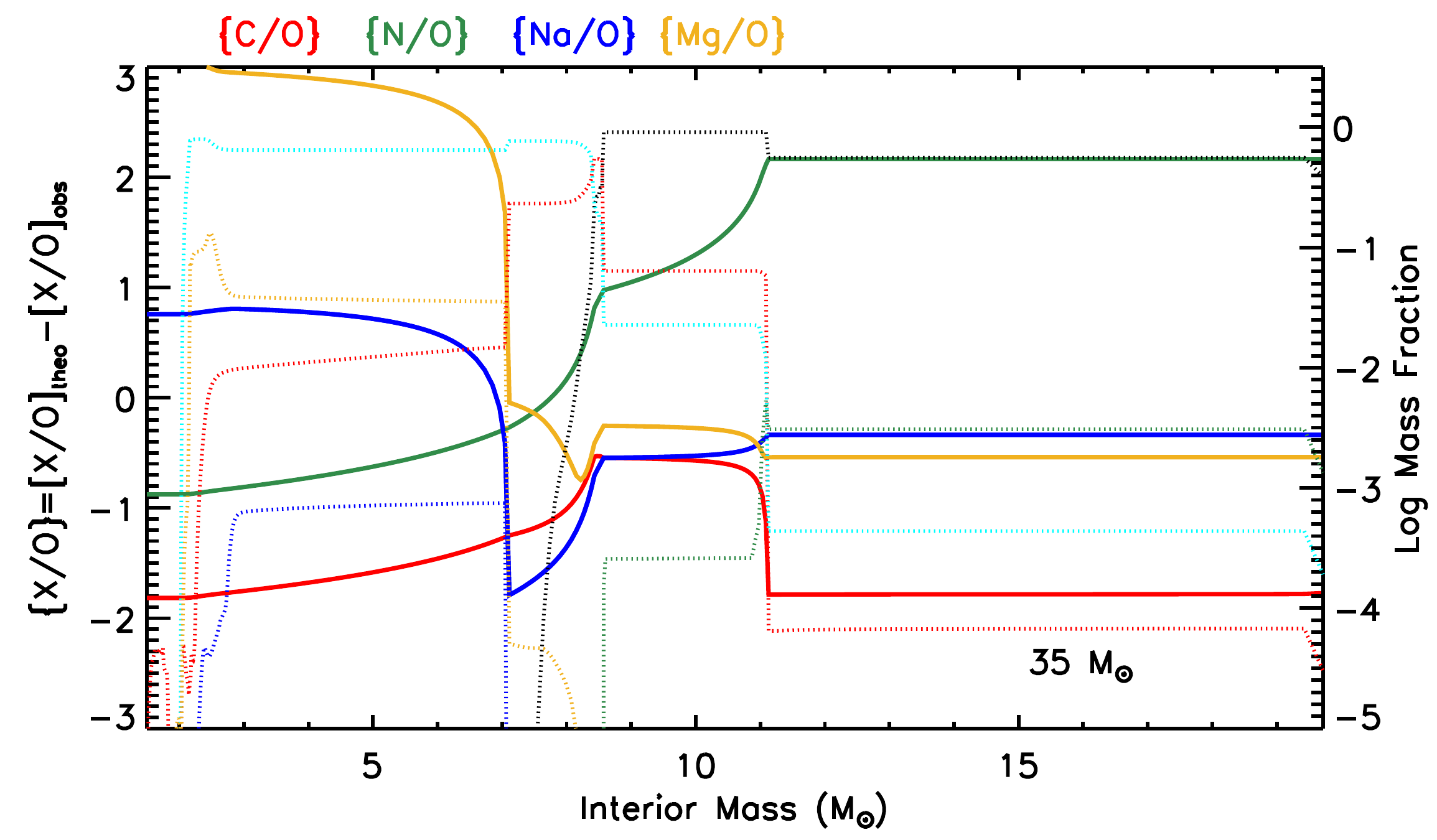}
\caption{Abundance ratios $\rm {C,N,Na,Mg/O}$ as a function of remnant mass for a zero-metallicity $\rm 15~M_\odot$  (top) and $\rm 35~M_\odot$ (bottom) star \citep{LC12} after the explosion (solid thick lines, left y-axis). The abundance ratio ${X/O}$ is defined as the difference between the predicted and observed [X/O] values. The right y-axis shows post-explosion interior profiles of $\rm ^{4}He$ (black dotted line), $\rm ^{12}C$ (red), $\rm ^{16}O$ (cyan), $\rm ^{23}Na$ (blue), and $\rm ^{24}Mg$ (orange).}
\label{fig:15msun}
\end{figure}

Subsequent spectroscopic studies of \ncstar\ allowed for a determination of the [O/Fe] abundance ratio (see Table\,\ref{tab:abbo} for literature values and Table\,\ref{tab:ourabbo} for our determination), which was unavailable at the time of the original analysis. The measured oxygen abundance — particularly the observed ratio $\rm [C/O] > 1$, implying a super-solar C/O ratio — presents a potential challenge for the model. This is because the only region in a massive star where $\rm ^{12}C$ exceeds $\rm ^{16}O$ in abundance is the He convective shell. In this zone, He burning is incomplete, leading to a significant accumulation of $\rm ^{12}C$ before it is fully converted into $\rm ^{16}O$.

As shown in the upper panel of Fig.\,\ref{fig:15msun}, a recent $\rm 15M_\odot$ zero-metallicity model \citep{LC12} exhibits a [C/O] ratio within the He convective shell (between mass coordinates $\sim 2.2$ and $\sim 3.2M_\odot$) that matches the observed value. Thus, placing the mass cut within this region yields ejecta with a [C/O] ratio in excellent agreement with observations. However, due to the initial zero metallicity, this region does not produce $\rm ^{14}N$, which is observed in \ncstar. Therefore, although such a mass cut can reproduce the observed [C/O], it fails to account for the observed [N/O].

As previously discussed, significant $\rm ^{14}N$  and $\rm ^{13}C$ production in zero-metallicity stars requires partial mixing between the He convective shell and the overlying H-rich shell. Proton ingestion into the He shell triggers a primary CNO cycle, generating $\rm ^{14}N$ and $\rm ^{13}C$ from freshly synthesised $\rm ^{12}C$. While this mixing process is relatively common in zero-metallicity stars, it occurs only in a limited mass range — typically between $\sim 25$ and $\sim 35~M_\odot$. Proton ingestion also alters the abundances of $\rm ^{12}C$ and $\rm ^{16}O$, reducing the [C/O] ratio below the observed value in \ncstar.

The bottom of Figure\,\ref{fig:15msun} shows the internal presupernova composition of a $\rm 35M_\odot$ model \citep{LC12} where such partial He/H mixing occurred. In the mixed region of the He shell (mass coordinates $\sim 11$ to $\sim 20M_\odot$), the [C/O] ratio drops significantly, while [N/O] rises above the observed value. Conversely, in the deeper portion of the He shell (between $\sim 8.5$ and $\sim 11~M_\odot$) that remained unmixed, $\rm ^{12}C$ dominates over $\rm ^{16}O$, resulting in a high [C/O] and low [N/O], along with [Na/O] and [Mg/O] values slightly below the observed ratios. These abundance patterns depend on the extent of He burning in the inner shell and the efficiency and depth of mixing with the H-rich envelope.

Among the models published by \citet{LC12}, this peculiar internal structure is found only in the $\rm 35~M_\odot$ case. However, due to the limited mass resolution of the available grid, it is currently not possible to fully explore how this behaviour varies with initial stellar mass. It remains plausible that models with intermediate or different initial masses could yield C, N, O, Na, and Mg abundance ratios more closely aligned with the observations. To investigate this possibility, we plan to conduct a more extensive and finely sampled study in future work.

\section{Conclusions}
The main conclusions are:
\begin{itemize}
    
\item {Observations of \ncstar\ over a longer temporal baseline have refined the orbital period to $10560^{+786}_{-518}$\,d ($\sim 28.9$\,y), which is smaller than previously derived \citep{aguado22}.}
\item {The binary system follows a halo prograde orbit around the Galactic centre.}
\item {New stringent upper limits on Sr and Ba abundances confirm that \ncstar\ satisfies the CEMP-no classification. These limits  do not support the existence  of a floor in the absolute barium abundance at $\rm A(Ba) \sim -2.0$ for CEMP-no stars, as proposed by \citet{thansen15}.}
\item {Lithium remains undetected at levels much lower than those reported by  \citet{mucciarelli22} for giants, implying that Li has been depleted either in the star itself or in its progenitor.}
\item {The Fe abundances derived under LTE assumptions from both \ion{Fe}{i} and \ion{Fe}{ii} lines are in good agreement, raising concerns about the reliability of published NLTE corrections for \ion{Fe}{i}.}
\item {The $\rm ^{12}C/^{13}C$ isotopic ratio is measured to be $110 \pm 12$, which is close to the solar value. We have shown that this ratio is sensitive to 3D NLTE effects, which tend to lower the derived value. The NLTE-corrected ratio is 87, higher than the average value observed in unmixed stars by \citet{spite06}, yet still low enough for $\rm ^{13}C$ to be the third most abundant element in the star’s atmosphere.}
\end{itemize}

\begin{acknowledgements}
PB acknowledges support   from the ERC advanced grant N. 835087 -- SPIAKID. 
LM gratefully acknowledges support from ANID-FONDECYT Regular Project n. 1251809.
This work has made use of data from the European Space Agency (ESA) mission
{\it Gaia} (\url{https://www.cosmos.esa.int/gaia}), processed by the {\it Gaia}
Data Processing and Analysis Consortium (DPAC,
\url{https://www.cosmos.esa.int/web/gaia/dpac/consortium}). Funding for the DPAC
has been provided by national institutions, in particular the institutions
participating in the {\it Gaia} Multilateral Agreement.
JIGH, DSA, RR and CAP acknowledge financial support from the Spanish Ministry of Science, Innovation and Universities (MICIU) projects PID2020-117493GB-I00 and PID2023-149982NB-I00.
This research has made use of the SIMBAD database, operated at CDS, Strasbourg, France.
PM warmly acknowledge Richard Hoppe for stressing the importance of the 3D/NLTE effects on the carbon isotopic ratio. NCS acknowledges funding by the European Union (ERC, FIERCE, 101052347). Views and opinions expressed are however those of the author(s) only and do not necessarily reflect those of the European Union or the European Research Council. Neither the European Union nor the granting authority can be held responsible for them. NJN was supported by FCT - Funda\c{c}\~ao para a Ci\^encia e a Tecnologia through national funds by grants UIDB/04434/2020 and UIDP/04434/2020.
The work of CJAPM was financed by Portuguese funds through FCT (Funda\c c\~ao para a Ci\^encia e a Tecnologia) in the framework of the project 2022.04048.PTDC (Phi in the Sky, DOI 10.54499/2022.04048.PTDC). CJM also acknowledges FCT and POCH/FSE (EC) support through Investigador FCT Contract 2021.01214.CEECIND/CP1658/CT0001 (DOI 10.54499/2021.01214.CEECIND/CP1658/CT0001). MTM acknowledges the support of the Australian Research Council through Future Fellowship grant FT180100194 and through the Australian Research Council Centre of Excellence in Optical Microcombs for Breakthrough Science (project number CE230100006) funded by the Australian Government.
\end{acknowledgements}

   \bibliographystyle{aa} 

   \bibliography{biblio}

\begin{thebibliography}{93}
\expandafter\ifx\csname natexlab\endcsname\relax\def\natexlab#1{#1}\fi

\bibitem[{{Aguado} {et~al.}(2022){Aguado}, {Molaro}, {Caffau}, {Gonz{\'a}lez
  Hern{\'a}ndez}, {Zapatero Osorio}, {Bonifacio}, {Allende Prieto}, {Rebolo},
  {Damasso}, {Su{\'a}rez Mascare{\~n}o}, {Howell}, {Furlan}, {Cristiani},
  {Cupani}, {Di Marcantonio}, {D'Odorico}, {Lovis}, {Martins}, {Milakovi},
  {Murphy}, {Nunes}, {Pepe}, {Santos}, {Schmidt}, \& {Sozzetti}}]{aguado22}
{Aguado}, D.~S., {Molaro}, P., {Caffau}, E., {et~al.} 2022, \aap, 668, A86

\bibitem[{{Allende Prieto} {et~al.}(2015){Allende Prieto},
  {Fern{\'a}ndez-Alvar}, {Aguado}, {Gonz{\'a}lez Hern{\'a}ndez}, {Rebolo},
  {Lee}, {Beers}, {Rockosi}, \& {Ge}}]{allendeprieto2015}
{Allende Prieto}, C., {Fern{\'a}ndez-Alvar}, E., {Aguado}, D.~S., {et~al.}
  2015, \aap, 579, A98

\bibitem[{{Amarsi} {et~al.}(2016){Amarsi}, {Lind}, {Asplund}, {Barklem}, \&
  {Collet}}]{amarsi2016}
{Amarsi}, A.~M., {Lind}, K., {Asplund}, M., {Barklem}, P.~S., \& {Collet}, R.
  2016, \mnras, 463, 1518

\bibitem[{{Andrievsky} {et~al.}(2011){Andrievsky}, {Spite}, {Korotin},
  {Fran{\c{c}}ois}, {Spite}, {Bonifacio}, {Cayrel}, \& {Hill}}]{andrievsky2011}
{Andrievsky}, S.~M., {Spite}, F., {Korotin}, S.~A., {et~al.} 2011, \aap, 530,
  A105

\bibitem[{{Andrievsky} {et~al.}(2010){Andrievsky}, {Spite}, {Korotin}, {Spite},
  {Bonifacio}, {Cayrel}, {Fran{\c{c}}ois}, \& {Hill}}]{andrievsky2010}
{Andrievsky}, S.~M., {Spite}, M., {Korotin}, S.~A., {et~al.} 2010, \aap, 509,
  A88

\bibitem[{{Andrievsky} {et~al.}(2007){Andrievsky}, {Spite}, {Korotin}, {Spite},
  {Bonifacio}, {Cayrel}, {Hill}, \& {Fran{\c{c}}ois}}]{andrievsky2007}
{Andrievsky}, S.~M., {Spite}, M., {Korotin}, S.~A., {et~al.} 2007, \aap, 464,
  1081

\bibitem[{{Andrievsky} {et~al.}(2008){Andrievsky}, {Spite}, {Korotin}, {Spite},
  {Bonifacio}, {Cayrel}, {Hill}, \& {Fran{\c{c}}ois}}]{andrievsky2008}
{Andrievsky}, S.~M., {Spite}, M., {Korotin}, S.~A., {et~al.} 2008, \aap, 481,
  481

\bibitem[{{Arentsen} {et~al.}(2019){Arentsen}, {Starkenburg}, {Shetrone},
  {Venn}, {Depagne}, \& {McConnachie}}]{arentsen19}
{Arentsen}, A., {Starkenburg}, E., {Shetrone}, M.~D., {et~al.} 2019, \aap, 621,
  A108

\bibitem[{{Beers} \& {Christlieb}(2005)}]{beers05}
{Beers}, T.~C. \& {Christlieb}, N. 2005, \araa, 43, 531

\bibitem[{{Beers} {et~al.}(1992){Beers}, {Preston}, \& {Shectman}}]{beers1992}
{Beers}, T.~C., {Preston}, G.~W., \& {Shectman}, S.~A. 1992, \aj, 103, 1987

\bibitem[{{Bensby} {et~al.}(2014){Bensby}, {Feltzing}, \& {Oey}}]{bensby14}
{Bensby}, T., {Feltzing}, S., \& {Oey}, M.~S. 2014, \aap, 562, A71

\bibitem[{{Bergemann} \& {Gehren}(2008)}]{bergemann08mn}
{Bergemann}, M. \& {Gehren}, T. 2008, \aap, 492, 823

\bibitem[{{Bessell} {et~al.}(2004){Bessell}, {Christlieb}, \&
  {Gustafsson}}]{bessel04}
{Bessell}, M.~S., {Christlieb}, N., \& {Gustafsson}, B. 2004, \apjl, 612, L61

\bibitem[{{Boesgaard} {et~al.}(2011){Boesgaard}, {Rich}, {Levesque}, \&
  {Bowler}}]{boesgaard2011}
{Boesgaard}, A.~M., {Rich}, J.~A., {Levesque}, E.~M., \& {Bowler}, B.~P. 2011,
  \apj, 743, 140

\bibitem[{{Bonifacio} {et~al.}(2018){Bonifacio}, {Caffau}, {Ludwig}, {Steffen},
  {Castelli}, {Gallagher}, {Ku{\v{c}}inskas}, {Prakapavi{\v{c}}ius}, {Cayrel},
  {Freytag}, {Plez}, \& {Homeier}}]{bonifacio2018}
{Bonifacio}, P., {Caffau}, E., {Ludwig}, H.~G., {et~al.} 2018, \aap, 611, A68

\bibitem[{{Bonifacio} {et~al.}(2015){Bonifacio}, {Caffau}, {Spite}, {Limongi},
  {Chieffi}, {Klessen}, {Fran{\c{c}}ois}, {Molaro}, {Ludwig}, {Zaggia},
  {Spite}, {Plez}, {Cayrel}, {Christlieb}, {Clark}, {Glover}, {Hammer}, {Koch},
  {Monaco}, {Sbordone}, \& {Steffen}}]{topos2}
{Bonifacio}, P., {Caffau}, E., {Spite}, M., {et~al.} 2015, \aap, 579, A28

\bibitem[{{Bonifacio} {et~al.}(2020){Bonifacio}, {Molaro}, {Adibekyan},
  {Aguado}, {Alibert}, {Allende Prieto}, {Caffau}, {Cristiani}, {Cupani}, {Di
  Marcantonio}, {D'Odorico}, {Ehrenreich}, {Figueira}, {Genova}, {Gonz{\'a}lez
  Hern{\'a}ndez}, {Lo Curto}, {Lovis}, {Martins}, {Mehner}, {Micela}, {Monaco},
  {Nunes}, {Pepe}, {Poretti}, {Rebolo}, {Santos}, {Saviane}, {Sousa},
  {Sozzetti}, {Suarez-Mascare{\~n}o}, {Udry}, \&
  {Zapatero-Osorio}}]{bonifacio20}
{Bonifacio}, P., {Molaro}, P., {Adibekyan}, V., {et~al.} 2020, \aap, 633, A129

\bibitem[{{Bonifacio} {et~al.}(1998){Bonifacio}, {Molaro}, {Beers}, \&
  {Vladilo}}]{bonifacio1998}
{Bonifacio}, P., {Molaro}, P., {Beers}, T.~C., \& {Vladilo}, G. 1998, \aap,
  332, 672

\bibitem[{{Bonifacio} {et~al.}(2021){Bonifacio}, {Monaco}, {Salvadori},
  {Caffau}, {Spite}, {Sbordone}, {Spite}, {Ludwig}, {Di Matteo}, {Haywood},
  {Fran{\c{c}}ois}, {Koch-Hansen}, {Christlieb}, \& {Zaggia}}]{bonifacio21}
{Bonifacio}, P., {Monaco}, L., {Salvadori}, S., {et~al.} 2021, \aap, 651, A79

\bibitem[{{Bovy}(2015)}]{bovy15}
{Bovy}, J. 2015, \apjs, 216, 29

\bibitem[{{Bovy} {et~al.}(2012){Bovy}, {Allende Prieto}, {Beers}, {Bizyaev},
  {da Costa}, {Cunha}, {Ebelke}, {Eisenstein}, {Frinchaboy}, {Garc{\'\i}a
  P{\'e}rez}, {Girardi}, {Hearty}, {Hogg}, {Holtzman}, {Maia}, {Majewski},
  {Malanushenko}, {Malanushenko}, {M{\'e}sz{\'a}ros}, {Nidever}, {O'Connell},
  {O'Donnell}, {Oravetz}, {Pan}, {Rocha-Pinto}, {Schiavon}, {Schneider},
  {Schultheis}, {Skrutskie}, {Smith}, {Weinberg}, {Wilson}, \&
  {Zasowski}}]{bovy12}
{Bovy}, J., {Allende Prieto}, C., {Beers}, T.~C., {et~al.} 2012, \apj, 759, 131

\bibitem[{{Caffau} {et~al.}(2011{\natexlab{a}}){Caffau}, {Bonifacio},
  {Fran{\c{c}}ois}, {Sbordone}, {Monaco}, {Spite}, {Spite}, {Ludwig}, {Cayrel},
  {Zaggia}, {Hammer}, {Randich}, {Molaro}, \& {Hill}}]{leo11}
{Caffau}, E., {Bonifacio}, P., {Fran{\c{c}}ois}, P., {et~al.}
  2011{\natexlab{a}}, \nat, 477, 67

\bibitem[{{Caffau} {et~al.}(2024{\natexlab{a}}){Caffau}, {Bonifacio}, {Monaco},
  {Sbordone}, {Spite}, {Fran{\c{c}}ois}, {Panuzzo}, {Sartoretti}, {Chemin},
  {Th{\'e}venin}, \& {Mucciarelli}}]{ghs143}
{Caffau}, E., {Bonifacio}, P., {Monaco}, L., {et~al.} 2024{\natexlab{a}}, \aap,
  684, L4

\bibitem[{{Caffau} {et~al.}(2024{\natexlab{b}}){Caffau}, {Bonifacio}, {Monaco},
  {Steffen}, {Sbordone}, {Spite}, {Fran{\c{c}}ois}, {Gallagher}, {Ludwig}, \&
  {Molaro}}]{leo2024}
{Caffau}, E., {Bonifacio}, P., {Monaco}, L., {et~al.} 2024{\natexlab{b}}, \aap,
  691, A245

\bibitem[{{Caffau} {et~al.}(2025){Caffau}, {Katz}, {Bonifacio}, {G{\'o}mez},
  {Lallement}, {Sartoretti}, {Royer}, {Panuzzo}, {Spite}, {Fran{\c{c}}ois},
  {Leclerc}, {Sbordone}, {Th{\'e}venin}, {Ludwig}, \& {Chemin}}]{rvs3}
{Caffau}, E., {Katz}, D., {Bonifacio}, P., {et~al.} 2025, Astronomische
  Nachrichten, 346, e70025

\bibitem[{{Caffau} {et~al.}(2011{\natexlab{b}}){Caffau}, {Ludwig}, {Steffen},
  {Freytag}, \& {Bonifacio}}]{caffau2011}
{Caffau}, E., {Ludwig}, H.~G., {Steffen}, M., {Freytag}, B., \& {Bonifacio}, P.
  2011{\natexlab{b}}, \solphys, 268, 255

\bibitem[{{Cayrel}(1988)}]{cayrel88}
{Cayrel}, R. 1988, in The Impact of Very High S/N Spectroscopy on Stellar
  Physics, ed. G.~{Cayrel de Strobel} \& M.~{Spite}, Vol. 132, 345

\bibitem[{{Cayrel} {et~al.}(2004){Cayrel}, {Depagne}, {Spite}, {Hill}, {Spite},
  {Fran{\c{c}}ois}, {Plez}, {Beers}, {Primas}, {Andersen}, {Barbuy},
  {Bonifacio}, {Molaro}, \& {Nordstr{\"o}m}}]{FS5}
{Cayrel}, R., {Depagne}, E., {Spite}, M., {et~al.} 2004, \aap, 416, 1117

\bibitem[{{Chamberlain} \& {Aller}(1951)}]{chamberlain51}
{Chamberlain}, J.~W. \& {Aller}, L.~H. 1951, \apj, 114, 52

\bibitem[{{Chieffi} {et~al.}(2001){Chieffi}, {Dom{\'\i}nguez}, {Limongi}, \&
  {Straniero}}]{chieffi+2001}
{Chieffi}, A., {Dom{\'\i}nguez}, I., {Limongi}, M., \& {Straniero}, O. 2001,
  \apj, 554, 1159

\bibitem[{{Chieffi} \& {Limongi}(2002)}]{CL02}
{Chieffi}, A. \& {Limongi}, M. 2002, \apj, 577, 281

\bibitem[{{Christlieb}(2008)}]{christlieb2008}
{Christlieb}, N. 2008, Physica Scripta Volume T, 133, 014034

\bibitem[{{Christlieb} {et~al.}(2002){Christlieb}, {Bessell}, {Beers},
  {Gustafsson}, {Korn}, {Barklem}, {Karlsson}, {Mizuno-Wiedner}, \&
  {Rossi}}]{christlieb02}
{Christlieb}, N., {Bessell}, M.~S., {Beers}, T.~C., {et~al.} 2002, \nat, 419,
  904

\bibitem[{{Christlieb} {et~al.}(2004){Christlieb}, {Gustafsson}, {Korn},
  {Barklem}, {Beers}, {Bessell}, {Karlsson}, \&
  {Mizuno-Wiedner}}]{christlieb04}
{Christlieb}, N., {Gustafsson}, B., {Korn}, A.~J., {et~al.} 2004, \apj, 603,
  708

\bibitem[{{Collet} {et~al.}(2006){Collet}, {Asplund}, \&
  {Trampedach}}]{collet06}
{Collet}, R., {Asplund}, M., \& {Trampedach}, R. 2006, \apjl, 644, L121

\bibitem[{{Collet} {et~al.}(2011){Collet}, {Hayek}, {Asplund}, {Nordlund},
  {Trampedach}, \& {Gudiksen}}]{collect2011}
{Collet}, R., {Hayek}, W., {Asplund}, M., {et~al.} 2011, \aap, 528, A32

\bibitem[{{Ezzeddine} {et~al.}(2017){Ezzeddine}, {Frebel}, \&
  {Plez}}]{ezzeddine17}
{Ezzeddine}, R., {Frebel}, A., \& {Plez}, B. 2017, \apj, 847, 142

\bibitem[{{Fernando} {et~al.}(2018){Fernando}, {Bernath}, {Hodges}, \&
  {Masseron}}]{fernando2018}
{Fernando}, A.~M., {Bernath}, P.~F., {Hodges}, J.~N., \& {Masseron}, T. 2018,
  \jqsrt, 217, 29

\bibitem[{{Feuillet} {et~al.}(2021){Feuillet}, {Sahlholdt}, {Feltzing}, \&
  {Casagrande}}]{feuillet21}
{Feuillet}, D.~K., {Sahlholdt}, C.~L., {Feltzing}, S., \& {Casagrande}, L.
  2021, \mnras, 508, 1489

\bibitem[{{Foreman-Mackey} {et~al.}(2017){Foreman-Mackey}, {Agol},
  {Ambikasaran}, \& {Angus}}]{celerite}
{Foreman-Mackey}, D., {Agol}, E., {Ambikasaran}, S., \& {Angus}, R. 2017, \aj,
  154, 220

\bibitem[{{Fran{\c{c}}ois} {et~al.}(2007){Fran{\c{c}}ois}, {Depagne}, {Hill},
  {Spite}, {Spite}, {Plez}, {Beers}, {Andersen}, {James}, {Barbuy}, {Cayrel},
  {Bonifacio}, {Molaro}, {Nordstr{\"o}m}, \& {Primas}}]{francois2007}
{Fran{\c{c}}ois}, P., {Depagne}, E., {Hill}, V., {et~al.} 2007, \aap, 476, 935

\bibitem[{{Freytag} {et~al.}(2012){Freytag}, {Steffen}, {Ludwig},
  {Wedemeyer-B{\"o}hm}, {Schaffenberger}, \& {Steiner}}]{freytag2012}
{Freytag}, B., {Steffen}, M., {Ludwig}, H.~G., {et~al.} 2012, Journal of
  Computational Physics, 231, 919

\bibitem[{{Fujimoto} {et~al.}(1990){Fujimoto}, {Iben}, \&
  {Hollowell}}]{fujimoto1990}
{Fujimoto}, M.~Y., {Iben}, Jr., I., \& {Hollowell}, D. 1990, \apj, 349, 580

\bibitem[{{Fulton} {et~al.}(2018){Fulton}, {Petigura}, {Blunt}, \&
  {Sinukoff}}]{fulton2018radvel}
{Fulton}, B.~J., {Petigura}, E.~A., {Blunt}, S., \& {Sinukoff}, E. 2018, \pasp,
  130, 044504

\bibitem[{{Gaia Collaboration} {et~al.}(2023){Gaia Collaboration}, {Vallenari},
  {Brown}, {Prusti}, {de Bruijne}, {Arenou}, {Babusiaux}, {Biermann},
  {Creevey}, {Ducourant}, {Evans}, {Eyer}, {Guerra}, {Hutton}, {Jordi},
  {Klioner}, {Lammers}, {Lindegren}, {Luri}, {Mignard}, {Panem}, {Pourbaix},
  {Randich}, {Sartoretti}, {Soubiran}, {Tanga}, {Walton}, {Bailer-Jones},
  {Bastian}, {Drimmel}, {Jansen}, {Katz}, {Lattanzi}, {van Leeuwen}, {Bakker},
  {Cacciari}, {Casta{\~n}eda}, {De Angeli}, {Fabricius}, {Fouesneau},
  {Fr{\'e}mat}, {Galluccio}, {Guerrier}, {Heiter}, {Masana}, {Messineo},
  {Mowlavi}, {Nicolas}, {Nienartowicz}, {Pailler}, {Panuzzo}, {Riclet}, {Roux},
  {Seabroke}, {Sordo}, {Th{\'e}venin}, {Gracia-Abril}, {Portell}, {Teyssier},
  {Altmann}, {Andrae}, {Audard}, {Bellas-Velidis}, {Benson}, {Berthier},
  {Blomme}, {Burgess}, {Busonero}, {Busso}, {C{\'a}novas}, {Carry}, {Cellino},
  {Cheek}, {Clementini}, {Damerdji}, {Davidson}, {de Teodoro}, {Nu{\~n}ez
  Campos}, {Delchambre}, {Dell'Oro}, {Esquej}, {Fern{\'a}ndez-Hern{\'a}ndez},
  {Fraile}, {Garabato}, {Garc{\'\i}a-Lario}, {Gosset}, {Haigron}, {Halbwachs},
  {Hambly}, {Harrison}, {Hern{\'a}ndez}, {Hestroffer}, {Hodgkin}, {Holl},
  {Jan{\ss}en}, {Jevardat de Fombelle}, {Jordan}, {Krone-Martins}, {Lanzafame},
  {L{\"o}ffler}, {Marchal}, {Marrese}, {Moitinho}, {Muinonen}, {Osborne},
  {Pancino}, {Pauwels}, {Recio-Blanco}, {Reyl{\'e}}, {Riello}, {Rimoldini},
  {Roegiers}, {Rybizki}, {Sarro}, {Siopis}, {Smith}, {Sozzetti}, {Utrilla},
  {van Leeuwen}, {Abbas}, {{\'A}brah{\'a}m}, {Abreu Aramburu}, {Aerts},
  {Aguado}, {Ajaj}, {Aldea-Montero}, {Altavilla}, {{\'A}lvarez}, {Alves},
  {Anders}, {Anderson}, {Anglada Varela}, {Antoja}, {Baines}, {Baker},
  {Balaguer-N{\'u}{\~n}ez}, {Balbinot}, {Balog}, {Barache}, {Barbato},
  {Barros}, {Barstow}, {Bartolom{\'e}}, {Bassilana}, {Bauchet}, {Becciani},
  {Bellazzini}, {Berihuete}, {Bernet}, {Bertone}, {Bianchi}, {Binnenfeld},
  {Blanco-Cuaresma}, {Blazere}, {Boch}, {Bombrun}, {Bossini}, {Bouquillon},
  {Bragaglia}, {Bramante}, {Breedt}, {Bressan}, {Brouillet}, {Brugaletta},
  {Bucciarelli}, {Burlacu}, {Butkevich}, {Buzzi}, {Caffau}, {Cancelliere},
  {Cantat-Gaudin}, {Carballo}, {Carlucci}, {Carnerero}, {Carrasco},
  {Casamiquela}, {Castellani}, {Castro-Ginard}, {Chaoul}, {Charlot}, {Chemin},
  {Chiaramida}, {Chiavassa}, {Chornay}, {Comoretto}, {Contursi}, {Cooper},
  {Cornez}, {Cowell}, {Crifo}, {Cropper}, {Crosta}, {Crowley}, {Dafonte},
  {Dapergolas}, {David}, {David}, {de Laverny}, {De Luise}, \& {De
  March}}]{gaiadr3}
{Gaia Collaboration}, {Vallenari}, A., {Brown}, A.~G.~A., {et~al.} 2023, \aap,
  674, A1

\bibitem[{{Gallagher} {et~al.}(2017){Gallagher}, {Caffau}, {Bonifacio},
  {Ludwig}, {Steffen}, {Homeier}, \& {Plez}}]{gallagher2017}
{Gallagher}, A.~J., {Caffau}, E., {Bonifacio}, P., {et~al.} 2017, \aap, 598,
  L10

\bibitem[{{Hansen} {et~al.}(2015){Hansen}, {Hansen}, {Christlieb}, {Beers},
  {Yong}, {Bessell}, {Frebel}, {Garc{\'\i}a P{\'e}rez}, {Placco}, {Norris}, \&
  {Asplund}}]{thansen15}
{Hansen}, T., {Hansen}, C.~J., {Christlieb}, N., {et~al.} 2015, \apj, 807, 173

\bibitem[{{Jedamzik} \& {Rehm}(2001)}]{jedamzik2001}
{Jedamzik}, K. \& {Rehm}, J.~B. 2001, \prd, 64, 023510

\bibitem[{{Kurucz}(2005)}]{K05}
{Kurucz}, R.~L. 2005, Memorie della Societa Astronomica Italiana Supplementi,
  8, 14

\bibitem[{{Ku{\v{c}}inskas} {et~al.}(2018){Ku{\v{c}}inskas}, {Klevas},
  {Ludwig}, {Bonifacio}, {Steffen}, \& {Caffau}}]{kucinskas2018}
{Ku{\v{c}}inskas}, A., {Klevas}, J., {Ludwig}, H.~G., {et~al.} 2018, \aap, 613,
  A24

\bibitem[{{Lardo} {et~al.}(2021){Lardo}, {Mashonkina}, {Jablonka}, {Bonifacio},
  {Caffau}, {Aguado}, {Gonz{\'a}lez Hern{\'a}ndez}, {Sestito}, {Kielty},
  {Venn}, {Hill}, {Starkenburg}, {Martin}, {Sitnova}, {Arentsen}, {Carlberg},
  {Navarro}, \& {Kordopatis}}]{PristineXIV}
{Lardo}, C., {Mashonkina}, L., {Jablonka}, P., {et~al.} 2021, \mnras, 508, 3068

\bibitem[{{Lebreton} \& {Reese}(2020)}]{LebretonReese2020}
{Lebreton}, Y. \& {Reese}, D.~R. 2020, \aap, 642, A88

\bibitem[{{Limberg} {et~al.}(2025){Limberg}, {Placco}, {Ji}, {Yao}, {Chiti},
  {Mardini}, {Frebel}, \& {Rossi}}]{limberg2025}
{Limberg}, G., {Placco}, V.~M., {Ji}, A.~P., {et~al.} 2025, \apjl, 989, L18

\bibitem[{{Limongi} \& {Chieffi}(2002)}]{LC02}
{Limongi}, M. \& {Chieffi}, A. 2002, \pasa, 19, 246

\bibitem[{{Limongi} \& {Chieffi}(2012)}]{LC12}
{Limongi}, M. \& {Chieffi}, A. 2012, \apjs, 199, 38

\bibitem[{{Limongi} {et~al.}(2003){Limongi}, {Chieffi}, \&
  {Bonifacio}}]{limongi+2003}
{Limongi}, M., {Chieffi}, A., \& {Bonifacio}, P. 2003, \apjl, 594, L123

\bibitem[{{Lindegren} {et~al.}(2021){Lindegren}, {Bastian}, {Biermann},
  {Bombrun}, {de Torres}, {Gerlach}, {Geyer}, {Hern{\'a}ndez}, {Hilger},
  {Hobbs}, {Klioner}, {Lammers}, {McMillan}, {Ramos-Lerate},
  {Steidelm{\"u}ller}, {Stephenson}, \& {van Leeuwen}}]{lindegren21}
{Lindegren}, L., {Bastian}, U., {Biermann}, M., {et~al.} 2021, \aap, 649, A4

\bibitem[{{Lodders} {et~al.}(2009){Lodders}, {Palme}, \& {Gail}}]{lodders2009}
{Lodders}, K., {Palme}, H., \& {Gail}, H.~P. 2009, Landolt B{\"o}rnstein, 4B,
  712

\bibitem[{{Ludwig} \& {Steffen}(2012)}]{ludwig2012}
{Ludwig}, H.-G. \& {Steffen}, M. 2012, in Astrophysics and Space Science
  Proceedings, Vol.~26, Red Giants as Probes of the Structure and Evolution of
  the Milky Way, ed. A.~{Miglio}, J.~{Montalb{\'a}n}, \& A.~{Noels}, 125

\bibitem[{{Ludwig} \& {Steffen}(2013)}]{ludwig2013}
{Ludwig}, H.~G. \& {Steffen}, M. 2013, Memorie della Societa Astronomica
  Italiana Supplementi, 24, 53

\bibitem[{{Malaney} \& {Fowler}(1989)}]{malaney1989}
{Malaney}, R.~A. \& {Fowler}, W.~A. 1989, \apjl, 345, L5

\bibitem[{{Manchon} {et~al.}(2024){Manchon}, {Deal}, {Goupil}, {Serenelli},
  {Lebreton}, {Klevas}, {Ku{\v{c}}inskas}, {Ludwig}, {Montalb{\'a}n}, \&
  {Gizon}}]{manchon2024}
{Manchon}, L., {Deal}, M., {Goupil}, M.~J., {et~al.} 2024, \aap, 687, A146

\bibitem[{{Matsuno} {et~al.}(2017){Matsuno}, {Aoki}, {Beers}, {Lee}, \&
  {Honda}}]{matsuno17}
{Matsuno}, T., {Aoki}, W., {Beers}, T.~C., {Lee}, Y.~S., \& {Honda}, S. 2017,
  \aj, 154, 52

\bibitem[{{Mel{\'e}ndez} \& {Barbuy}(2009)}]{melendez2009}
{Mel{\'e}ndez}, J. \& {Barbuy}, B. 2009, \aap, 497, 611

\bibitem[{{Molaro} {et~al.}(2023){Molaro}, {Aguado}, {Caffau}, {Allende
  Prieto}, {Bonifacio}, {Gonz{\'a}lez Hern{\'a}ndez}, {Rebolo}, {Zapatero
  Osorio}, {Cristiani}, {Pepe}, {Santos}, {Alibert}, {Cupani}, {Di
  Marcantonio}, {D'Odorico}, {Lovis}, {Martins}, {Milakovi{\'c}}, {Murphy},
  {Nunes}, {Schmidt}, {Sousa}, {Sozzetti}, \& {Su{\'a}rez
  Mascare{\~n}o}}]{molaro23}
{Molaro}, P., {Aguado}, D.~S., {Caffau}, E., {et~al.} 2023, \aap, 679, A72

\bibitem[{{Mucciarelli} {et~al.}(2022){Mucciarelli}, {Monaco}, {Bonifacio},
  {Salaris}, {Deal}, {Spite}, {Richard}, \& {Lallement}}]{mucciarelli22}
{Mucciarelli}, A., {Monaco}, L., {Bonifacio}, P., {et~al.} 2022, \aap, 661,
  A153

\bibitem[{{Nakamura} \& {Hashimoto}(2017)}]{nakamura2017}
{Nakamura}, R. \& {Hashimoto}, M.-a. 2017, in 14th International Symposium on
  Nuclei in the Cosmos (NIC2016), ed. S.~{Kubono}, T.~{Kajino}, S.~{Nishimura},
  T.~{Isobe}, S.~{Nagataki}, T.~{Shima}, \& Y.~{Takeda}, 020108

\bibitem[{{Nave} \& {Johansson}(2013)}]{nave2013}
{Nave}, G. \& {Johansson}, S. 2013, \apjs, 204, 1

\bibitem[{{Nordlander} {et~al.}(2017){Nordlander}, {Amarsi}, {Lind}, {Asplund},
  {Barklem}, {Casey}, {Collet}, \& {Leenaarts}}]{nordlander2017}
{Nordlander}, T., {Amarsi}, A.~M., {Lind}, K., {et~al.} 2017, \aap, 597, A6

\bibitem[{{Nordlander} {et~al.}(2019){Nordlander}, {Bessell}, {Da Costa},
  {Mackey}, {Asplund}, {Casey}, {Chiti}, {Ezzeddine}, {Frebel}, {Lind},
  {Marino}, {Murphy}, {Norris}, {Schmidt}, \& {Yong}}]{nordlander2019}
{Nordlander}, T., {Bessell}, M.~S., {Da Costa}, G.~S., {et~al.} 2019, \mnras,
  488, L109

\bibitem[{{Norris} {et~al.}(1997){Norris}, {Ryan}, \& {Beers}}]{norris1997}
{Norris}, J.~E., {Ryan}, S.~G., \& {Beers}, T.~C. 1997, \apjl, 489, L169

\bibitem[{{Pietrinferni} {et~al.}(2006){Pietrinferni}, {Cassisi}, {Salaris}, \&
  {Castelli}}]{basti}
{Pietrinferni}, A., {Cassisi}, S., {Salaris}, M., \& {Castelli}, F. 2006, \apj,
  642, 797

\bibitem[{{Pietrinferni} {et~al.}(2021{\natexlab{a}}){Pietrinferni}, {Hidalgo},
  {Cassisi}, {Salaris}, {Savino}, {Mucciarelli}, {Verma}, {Silva Aguirre},
  {Aparicio}, \& {Ferguson}}]{pietrinferni2021}
{Pietrinferni}, A., {Hidalgo}, S., {Cassisi}, S., {et~al.} 2021{\natexlab{a}},
  \apj, 908, 102

\bibitem[{{Pietrinferni} {et~al.}(2021{\natexlab{b}}){Pietrinferni}, {Hidalgo},
  {Cassisi}, {Salaris}, {Savino}, {Mucciarelli}, {Verma}, {Silva Aguirre},
  {Aparicio}, \& {Ferguson}}]{Pietrinferni_2021}
{Pietrinferni}, A., {Hidalgo}, S., {Cassisi}, S., {et~al.} 2021{\natexlab{b}},
  \apj, 908, 102

\bibitem[{{Planck Collaboration} {et~al.}(2020){Planck Collaboration},
  {Aghanim}, {Akrami}, {Ashdown}, {Aumont}, {Baccigalupi}, {Ballardini},
  {Banday}, {Barreiro}, {Bartolo}, {Basak}, {Battye}, {Benabed}, {Bernard},
  {Bersanelli}, {Bielewicz}, {Bock}, {Bond}, {Borrill}, {Bouchet}, {Boulanger},
  {Bucher}, {Burigana}, {Butler}, {Calabrese}, {Cardoso}, {Carron},
  {Challinor}, {Chiang}, {Chluba}, {Colombo}, {Combet}, {Contreras}, {Crill},
  {Cuttaia}, {de Bernardis}, {de Zotti}, {Delabrouille}, {Delouis}, {Di
  Valentino}, {Diego}, {Dor{\'e}}, {Douspis}, {Ducout}, {Dupac}, {Dusini},
  {Efstathiou}, {Elsner}, {En{\ss}lin}, {Eriksen}, {Fantaye}, {Farhang},
  {Fergusson}, {Fernandez-Cobos}, {Finelli}, {Forastieri}, {Frailis},
  {Fraisse}, {Franceschi}, {Frolov}, {Galeotta}, {Galli}, {Ganga},
  {G{\'e}nova-Santos}, {Gerbino}, {Ghosh}, {Gonz{\'a}lez-Nuevo}, {G{\'o}rski},
  {Gratton}, {Gruppuso}, {Gudmundsson}, {Hamann}, {Handley}, {Hansen},
  {Herranz}, {Hildebrandt}, {Hivon}, {Huang}, {Jaffe}, {Jones}, {Karakci},
  {Keih{\"a}nen}, {Keskitalo}, {Kiiveri}, {Kim}, {Kisner}, {Knox},
  {Krachmalnicoff}, {Kunz}, {Kurki-Suonio}, {Lagache}, {Lamarre}, {Lasenby},
  {Lattanzi}, {Lawrence}, {Le Jeune}, {Lemos}, {Lesgourgues}, {Levrier},
  {Lewis}, {Liguori}, {Lilje}, {Lilley}, {Lindholm}, {L{\'o}pez-Caniego},
  {Lubin}, {Ma}, {Mac{\'\i}as-P{\'e}rez}, {Maggio}, {Maino}, {Mandolesi},
  {Mangilli}, {Marcos-Caballero}, {Maris}, {Martin}, {Martinelli},
  {Mart{\'\i}nez-Gonz{\'a}lez}, {Matarrese}, {Mauri}, {McEwen}, {Meinhold},
  {Melchiorri}, {Mennella}, {Migliaccio}, {Millea}, {Mitra},
  {Miville-Desch{\^e}nes}, {Molinari}, {Montier}, {Morgante}, {Moss}, {Natoli},
  {N{\o}rgaard-Nielsen}, {Pagano}, {Paoletti}, {Partridge}, {Patanchon},
  {Peiris}, {Perrotta}, {Pettorino}, {Piacentini}, {Polastri}, {Polenta},
  {Puget}, {Rachen}, {Reinecke}, {Remazeilles}, {Renzi}, {Rocha}, {Rosset},
  {Roudier}, {Rubi{\~n}o-Mart{\'\i}n}, {Ruiz-Granados}, {Salvati}, {Sandri},
  {Savelainen}, {Scott}, {Shellard}, {Sirignano}, {Sirri}, {Spencer},
  {Sunyaev}, {Suur-Uski}, {Tauber}, {Tavagnacco}, {Tenti}, {Toffolatti},
  {Tomasi}, {Trombetti}, {Valenziano}, {Valiviita}, {Van Tent}, {Vibert},
  {Vielva}, {Villa}, {Vittorio}, {Wandelt}, {Wehus}, {White}, {White},
  {Zacchei}, \& {Zonca}}]{planck}
{Planck Collaboration}, {Aghanim}, N., {Akrami}, Y., {et~al.} 2020, \aap, 641,
  A6

\bibitem[{{Prakapavi{\v{c}}ius} {et~al.}(2017){Prakapavi{\v{c}}ius},
  {Ku{\v{c}}inskas}, {Dobrovolskas}, {Klevas}, {Steffen}, {Bonifacio},
  {Ludwig}, \& {Spite}}]{prakapavicius2017}
{Prakapavi{\v{c}}ius}, D., {Ku{\v{c}}inskas}, A., {Dobrovolskas}, V., {et~al.}
  2017, \aap, 599, A128

\bibitem[{{Reese} \& {Lebreton}(2020)}]{spins2020}
{Reese}, D.~R. \& {Lebreton}, Y. 2020, {SPInS: Stellar Parameters INferred
  Systematically}, Astrophysics Source Code Library, record ascl:2009.006

\bibitem[{{Rizzuti} {et~al.}(2025){Rizzuti}, {Cescutti}, {Molaro}, {Roberti},
  {Chieffi}, {Limongi}, {Magrini}, \& {Matteucci}}]{rizzuti2025}
{Rizzuti}, F., {Cescutti}, G., {Molaro}, P., {et~al.} 2025, \aap, 698, A118

\bibitem[{{Roberti} {et~al.}(2024){Roberti}, {Limongi}, \&
  {Chieffi}}]{roberti+2024}
{Roberti}, L., {Limongi}, M., \& {Chieffi}, A. 2024, \apjs, 270, 28

\bibitem[{{Schlafly} \& {Finkbeiner}(2011)}]{schlafly11}
{Schlafly}, E.~F. \& {Finkbeiner}, D.~P. 2011, \apj, 737, 103

\bibitem[{{Schnabel} {et~al.}(2004){Schnabel}, {Schultz-Johanning}, \&
  {Kock}}]{schnabel2004}
{Schnabel}, R., {Schultz-Johanning}, M., \& {Kock}, M. 2004, \aap, 414, 1169

\bibitem[{{Sch{\"o}nrich} {et~al.}(2010){Sch{\"o}nrich}, {Binney}, \&
  {Dehnen}}]{schonrich10}
{Sch{\"o}nrich}, R., {Binney}, J., \& {Dehnen}, W. 2010, \mnras, 403, 1829

\bibitem[{{Sestito} {et~al.}(2019){Sestito}, {Longeard}, {Martin},
  {Starkenburg}, {Fouesneau}, {Gonz{\'a}lez Hern{\'a}ndez}, {Arentsen},
  {Ibata}, {Aguado}, {Carlberg}, {Jablonka}, {Navarro}, {Tolstoy}, \&
  {Venn}}]{sestito19}
{Sestito}, F., {Longeard}, N., {Martin}, N.~F., {et~al.} 2019, \mnras, 484,
  2166

\bibitem[{{Sitnova} {et~al.}(2019){Sitnova}, {Mashonkina}, {Ezzeddine}, \&
  {Frebel}}]{sitnova19}
{Sitnova}, T.~M., {Mashonkina}, L.~I., {Ezzeddine}, R., \& {Frebel}, A. 2019,
  \mnras, 485, 3527

\bibitem[{{Smiljanic} {et~al.}(2009){Smiljanic}, {Pasquini}, {Bonifacio},
  {Galli}, {Gratton}, {Randich}, \& {Wolff}}]{smiljanic2009}
{Smiljanic}, R., {Pasquini}, L., {Bonifacio}, P., {et~al.} 2009, \aap, 499, 103

\bibitem[{{Speagle}(2020)}]{speagle20dynesty}
{Speagle}, J.~S. 2020, \mnras, 493, 3132

\bibitem[{{Spite} {et~al.}(2019){Spite}, {Bonifacio}, {Spite}, {Caffau},
  {Sbordone}, \& {Gallagher}}]{spite2019}
{Spite}, M., {Bonifacio}, P., {Spite}, F., {et~al.} 2019, \aap, 624, A44

\bibitem[{{Spite} {et~al.}(2006){Spite}, {Cayrel}, {Hill}, {Spite},
  {Fran{\c{c}}ois}, {Plez}, {Bonifacio}, {Molaro}, {Depagne}, {Andersen},
  {Barbuy}, {Beers}, {Nordstr{\"o}m}, \& {Primas}}]{spite06}
{Spite}, M., {Cayrel}, R., {Hill}, V., {et~al.} 2006, \aap, 455, 291

\bibitem[{{Starkenburg} {et~al.}(2018){Starkenburg}, {Aguado}, {Bonifacio},
  {Caffau}, {Jablonka}, {Lardo}, {Martin}, {S{\'a}nchez-Janssen}, {Sestito},
  {Venn}, {Youakim}, {Allende Prieto}, {Arentsen}, {Gentile}, {Gonz{\'a}lez
  Hern{\'a}ndez}, {Kielty}, {Koppelman}, {Longeard}, {Tolstoy}, {Carlberg},
  {C{\^o}t{\'e}}, {Fouesneau}, {Hill}, {McConnachie}, \&
  {Navarro}}]{starkenburg18}
{Starkenburg}, E., {Aguado}, D.~S., {Bonifacio}, P., {et~al.} 2018, \mnras,
  481, 3838

\bibitem[{{Takeda} {et~al.}(2005){Takeda}, {Hashimoto}, {Taguchi}, {Yoshioka},
  {Takada-Hidai}, {Saito}, \& {Honda}}]{takeda2005}
{Takeda}, Y., {Hashimoto}, O., {Taguchi}, H., {et~al.} 2005, \pasj, 57, 751

\bibitem[{{Tauris} \& {van den Heuvel}(2006)}]{TvdH06}
{Tauris}, T.~M. \& {van den Heuvel}, E.~P.~J. 2006, in Compact stellar X-ray
  sources, Vol.~39, 623--665

\bibitem[{{Thomas} {et~al.}(1994){Thomas}, {Schramm}, {Olive}, {Mathews},
  {Meyer}, \& {Fields}}]{thomas1994}
{Thomas}, D., {Schramm}, D.~N., {Olive}, K.~A., {et~al.} 1994, \apj, 430, 291

\bibitem[{{Yong} {et~al.}(2013){Yong}, {Norris}, {Bessell}, {Christlieb},
  {Asplund}, {Beers}, {Barklem}, {Frebel}, \& {Ryan}}]{yong13}
{Yong}, D., {Norris}, J.~E., {Bessell}, M.~S., {et~al.} 2013, \apj, 762, 26

\end{thebibliography}

\begin{appendix}

\section{Data from the literature} \label{sec:litdata}

The abundances and upper limits derived in the literture are reported in Table\,\ref{tab:abbo}, while in Table\,\ref{tab:ourabbo} our results are listed.

\begin{table*}
\caption{Abundances from the literature.}
\label{tab:abbo}
\footnotesize\setlength{\extrarowheight}{2pt}
\begin{tabular}{l|r|r|r|r|r|r|r|r|r}
\hline
\smallskip
Element&  NC02 & NC04 & B04 & RC06                    & Y13 & E17 &S19 & A22 & M23\\
       & [X/Fe]& A(X) &     &                         &     &     & C-NLTE$^1$     &     &    \\
\hline 
\Teff   & 5100 & 5100 & 5100 & 5130 & 5100 & 5050 & 5300 & 5100 & 5100 \\
\logg   & 2.2  & 2.2  & 2.2  & 2.2  & 2.2  & 2.3  & 2.5  & 2.2  & 2.2 \\ 
Li     & $< 5.3       $         & $   < 1.12     $     &                 &  $< 0.94                    $ &                 &               &                  &  $< 0.50$  &   \\
C      & $  4.0 $              & $6.81/7.11^2   $   &                 &    $\sim 5.7^3$      & $ 6.74$         &               &                  &  $  6.90$  &  $  6.75$ \\
12C/13C& $            $         & $\sim 60       $     &                 &  $                          $ &                 &               &                  &  $  87  $  &   \\
N      & $  2.3       $         & $     4.93/5.22^{4}$     &                 &  $\sim 3^5     $ & $ 4.72$         &               &                  &  $  5.00$  &  $      $ \\
O      & $            $         & $              $     &   5.66          &  $\sim 5^6$ &                 &               &                  &  $      $  &  $      $ \\
Na     & $  0.8       $         & $     1.86     $     &                 &  $  1.75                    $ & $ 1.81$         &               &                  &  $  1.62$  &  $      $ \\  
Mg     & $  0.2       $         & $     2.41     $     &                 &  $  2.33                    $ & $ 2.32$         &               &  0.30            &  $  2.24$  &  $      $ \\
Al     & $            $         & $   < 0.93     $     &                 &  $< 0.79                    $ &                 &               &                  &  $< 0.90$  &  $      $ \\
Si     & $            $         & $   < 2.55     $     &                 &  $< 2.45                    $ &                 &               &                  &  $      $  &  $      $ \\
S      & $            $         & $   < 7.11     $     &                 &  $< 7.15                    $ &                 &               &                  &  $      $  &  $      $ \\
K      & $            $         & $              $     &                 &  $                          $ &                 &               &                  &  $< 1.22$  &  $      $ \\
CaI    & $  0.4       $         & $     0.99     $     &                 &  $  0.84                    $ & $ 0.92$         &               &  0.02            &  $  0.78$  &  $      $ \\   
CaII   & $            $         & $     1.44     $     &                 &  $  1.24                    $ &                 &               &                  &  $  1.43$  &  $      $ \\
Sc   & $            $         & $   <-1.50     $     &                 &  $<-1.63                    $ &                 &               &                  &  $      $  &  $      $ \\
Ti     & $ -0.4       $         & $    -0.62     $     &                 &  $ -0.66                    $ & $-0.55$         &               &                  &  $      $  &  $      $ \\
Cr     & $            $         & $   < 0.65     $     &                 &  $< 0.36                    $ &                 &               &                  &  $< 0.67$  &  $      $ \\  
Mn     & $            $         & $   < 0.47     $     &                 &  $< 0.08                    $ &                 &               &                  &  $      $  &  $      $ \\
FeI     & $ -5.3\pm 0.2$         & $2.06/2.17^7     $  &                 &  $  1.83                    $ & $ 1.96$         & $2.03\pm 0.20/2.78\pm 0.15^7$&                  &  $  1.96$  &           \\
FeII   & $            $         & $   < 3.00     $     &                 &  $< 3.06                    $ &                 &               &                  &  $      $  &  $      $ \\
Co     & $            $         & $   < 0.86     $     &                 &  $< 0.60                    $ &                 &               &                  &  $      $  &  $      $ \\
Ni     & $ -0.4       $         & $     0.60     $     &                 &  $  0.27                    $ &                 &               &                  &  $      $  &  $      $ \\
Zn     & $< 2.7       $         & $   < 1.97     $     &                 &  $< 2.02                    $ &                 &               &                  &  $      $  &  $      $ \\
Sr     & $<-0.5       $         & $   <-2.83     $     &                 &  $<-3.00                    $ & $<-3.19$        &               &                  &  $<-3.40$  &   \\
Ba     & $< 0.8       $         & $   <-2.33     $     &                 &  $<-2.59                    $ & $<-2.54$        &               &                  &  $<-2.87$  &  $<-3.19$ \\
Eu     & $< 2.8       $         & $   <-1.99     $     &                 &  $<-2.52                    $ &                 &               &                  &  $      $     $      $ \\  
\hline
\end{tabular}
\\
NC02: \citet{christlieb02}; 
NC04: \citet{christlieb04};
B04: \citet{bessel04};
RC06: \citet{collet06};
Y13: \citet{yong13};
E17: \citet{ezzeddine17};
S19: \citet{sitnova19};
A22: \citet{aguado22};
M23: \citet{molaro23}.\\
$^1$ NLTE corrections.\\
$^2$ From CH and C2 lines, respectively.\\
$^3$ average assuming two different A(C) of 7.11 and 6.81, respectively and on different lines.\\
$^4$ consistent values from CH and C2 lines.\\
$^5$ from CN lines.\\
$^6$ average value after 3D correction of several synthetic OH lines.\\
$^7$ LTE and NLTE.\\
\end{table*}

\begin{table}
\caption{Abundances derived in this work.}
\label{tab:ourabbo}
\footnotesize\setlength{\extrarowheight}{2pt}
\begin{tabular}{lrrrrr}
\hline
\smallskip
Element& $\rm A(X)_\odot$ & A(X) & [X/H] & [X/Fe] & 3D LTE corr \\
\hline 
\smallskip
\ion{Fe}{i}$^{(*)}$ & $7.52$ & $1.96$   & $-5.56$  & \\ 
\ion{Be}{i}  & $1.38$ & $<-2.05$ & $<-3.44$ & $<2.12$  \\
\ion{Li}{i}  & 1.03   & $<-0.14$ &          &          \\
\ion{$^{12}$C}{i} &        & $ 6.72$ &          &          &$-0.65$\\
\ion{$^{13}$C}{i} &        & $ 4.68$ &          &          \\
\ion{N}{i}   & $7.86$ & $ 4.42$  & $-3.44$  & $+2.12$& $-0.58$ \\
\ion{O}{i}   & $8.76$ & $ 5.72$  & $-3.04$  & $+2.52$& $-0.58$ \\
\ion{Na}{i}  & $6.30$ & $1.64$   & $-4.66$  & $0.90$ \\
\ion{Al}{i}  & $6.47$ & $0.27:$  & $-6.20:$ & $-0.64:$ \\
\ion{Si}{i}  & $7.52$ & $<1.8$   & $<-5.72$ & $<-0.16$ \\
\ion{S}{i}   & $7.16$ & $<4.40$  & $<-2.76$ & $<2.80$  \\
\ion{K}{i}   & $5.11$ & $<1.05$  & $<-4.06$ & $<1.56$  \\
\ion{Sc}{ii} & $3.10$ & $-2.30$  & $-5.40$ & $0.16$  \\
\ion{Cr}{i}  & $5.64$ & $<0.30$  & $<-5.34$ & $<0.22$  \\
\ion{Mn}{i}  & $5.26$ & $<-0.50$  & $<-5.76$ & $<-0.20$ \\
\ion{Fe}{ii} & $7.52$ & $1.74$  &          &   & $+0.10$\\ 
\ion{Co}{i}  & $4.92$ & $0.30:$   &  $-4.62:$ & $+0.94:$ \\
\ion{Zn}{i}  & $4.62$ & $<1.02$  & $<-3.60$ & $<1.96$  \\ 
\ion{Sr}{ii} & $2.92$ & $<-3.89$  & $<-6.81$ & $<-1.25$ \\
\ion{Ba}{ii} & $2.17$ & $<-3.87$  & $<-6.04$ & $<-0.48$ \\
\ion{Eu}{ii} & $0.51$ & $<-2.67$ & $<-3.18$ & $<2.38$  \\
\hline
\end{tabular}
\\
$^{(*)}$ A(Fe) is from \citet{aguado22} and it is used in the table for [Fe/H] to compute [X/Fe] for all the elements.
\end{table}

\section{Model atmospheres and spectrum synthesis} \label{sec:models}
\subsection{1D (LTE) models and spectrum synthesis} \label{sec:A1}
As 1D model atmosphere we computed an ATLAS\,9 \citep{K05}
model with \teff = 5100\,K and \logg\, 2.2,
with metallicity [M/H]=$-5.0$, $\alpha$ enhancement +0.4\,dex and microturbulence $1.0$\,\kms,
using the Opacity Distribution Function of Mucciarelli et al. (in preparation). The effect of microturbulence on the model structure and emerging flux is minor \citep[see][]{kucinskas2018}.
The ODF used assumes the (scaled) solar abundances of \citet{caffau2011} and
\citet{lodders2009} for the elements not included in the former paper.
This is also the set of solar abundances assumed elsewhere in this work.

We also computed an ATLAS\,12 model atmosphere with the precise chemical composition
of \ncstar . We verified that the ATLAS\,9 and ATLAS\,12 models
have exactly the same temperature structure for all layers
that are relevant for the line formation. The two models begin to differ
for $\log(\tau_{\rm Ross}) < -4$, the ATLAS\,12 model being cooler. This difference is not due
to the different chemical composition but is due to the different temperature
correction algorithm in the two codes.
One should bear in mind that real stars are expected to show a chromospheric temperature rise
at very low optical depths that cannot be modelled by the 1D LTE approach of the 
ATLAS codes. We typically compute models up to optical depth of
$\log(\tau_{\rm Ross}) \sim -6.5$ just to avoid an edge effect when
computing the cores of strong lines. We do not claim that
these are correctly computed, since they certainly are sensitive
to the structure of the chromosphere and NLTE effects.

When computing the ATLAS models we used the default control card
{\tt SCATTERING ON} that includes the scattering
term in the computation of the continuum source function.
The effect of scattering is discussed in \citet[][section 2.3]{bonifacio2018}.

To compute the synthetic spectra from our 1D model atmosphere we used the SYNTHE 
suite \citep{K05}. The scattering is treated explicitly and
without approximation. In fact, this is specially important in the UV range: if
one does not treat scattering explicitly one finds a trend of abundances with 
wavelength \citep[see][section 3.2 for a discussion]{FS5}.

\subsection{3D (LTE) models and abundance corrections} \label{sec:A2}
Two dedicated high-resolution CO5BOLD models ($288\times 288\times 200$ grid)
have been computed to represent the dynamic atmosphere of HE\,0107--5240
($T_{\rm eff}\sim 5100$\,K, $\log g=2.2$). For the first model, we assumed
[M/H]=$-4.0$ solar scaled abundances (with an $\alpha$ enhancement of
$+0.4$\,dex), for the second one an enhancement of C and N by $3.0$\,dex and
of O by $2.6$\,dex was applied with respect to the first model. The detailed chemical
composition is relevant for constructing the 14-band opacity tables used for
computing the radiation transport in CO5BOLD. Continuum scattering is treated
approximately by ignoring the scattering opacity in the optically thin layers
as suggested by \citet{collect2011}, \citep[see also][]{ludwig2012,ludwig2013}.

\begin{figure}
\centering
\includegraphics[width=\hsize,clip=true]{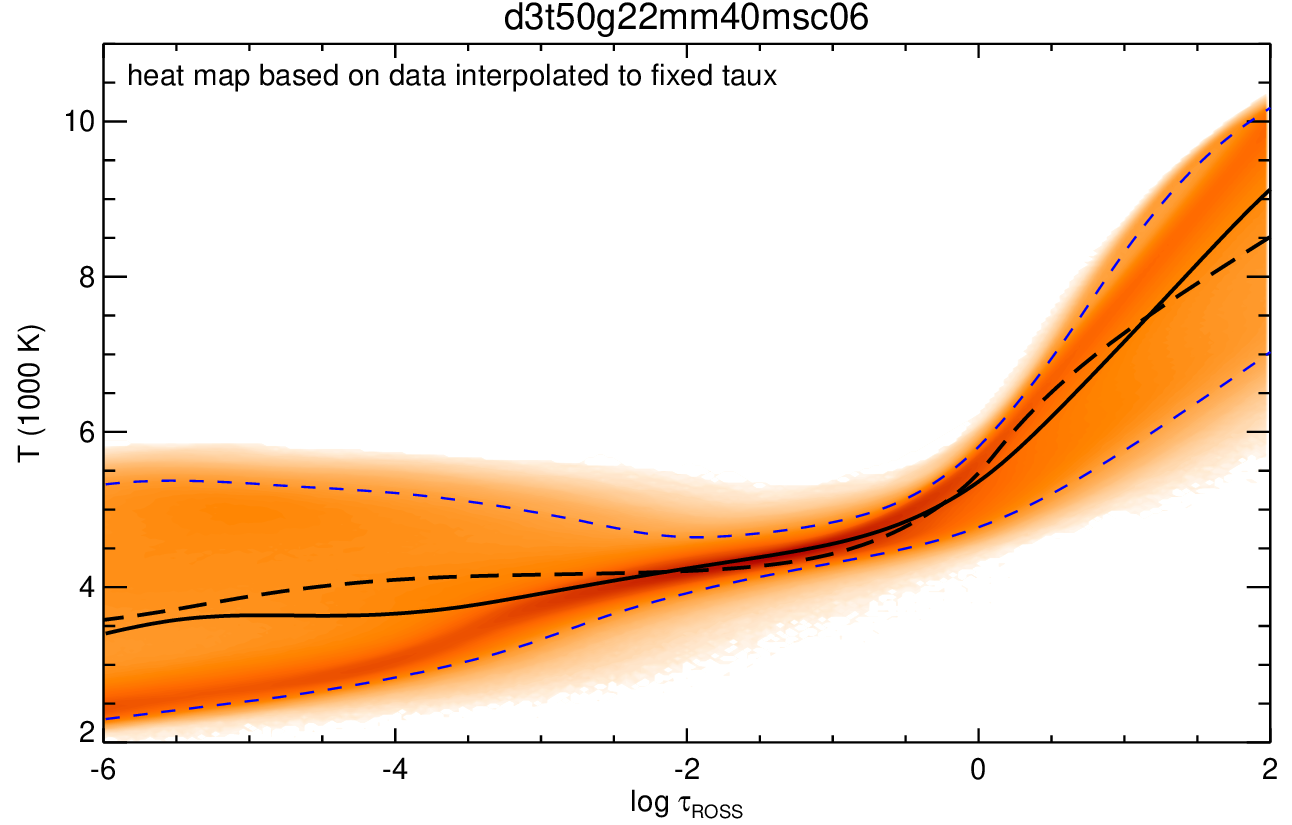}\\[3mm]
\includegraphics[width=\hsize,clip=true]{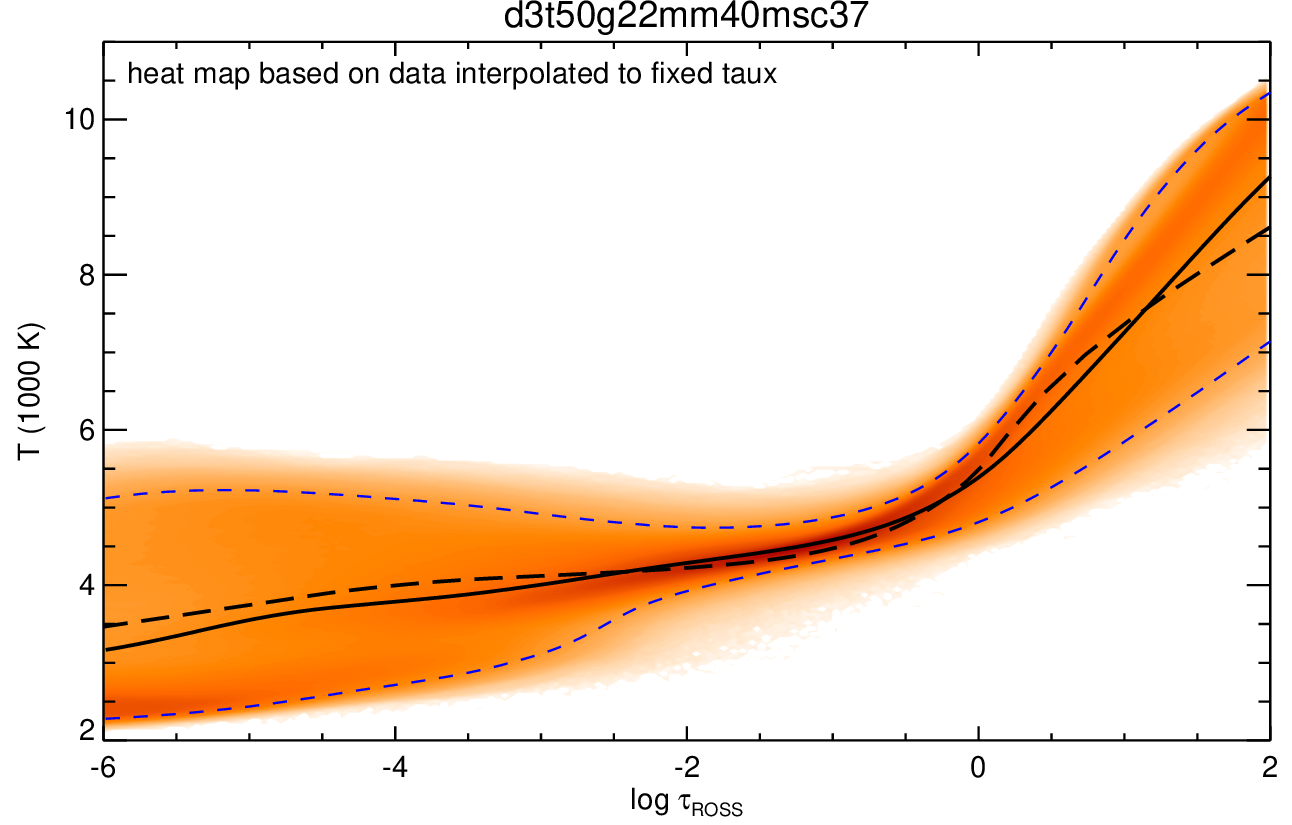}
\caption{
3D versus 1D temperature structure of two 3D CO5BOLD models representing
  HE\,0107--5240 ($T_{\rm eff}\sim 5090$\,K, $\log g=2.2$, [M/H]=-4.0, $\alpha$
  enhancement $+0.4$), assuming different CNO abundances for computing the
  opacity tables:
  \emph{Top}: A(C)=4.39, A (N)=3.78, A(O)=5.06;
  \emph{Bottom}: A(C)=7.39, A(N)=6.78, A(O)=7.66. 
  The orange band outlines the 3D temperature distribution constructed from $20$
  snapshots of the respective high-resolution model, each resampled to
  $144\times 144\times 200$ grid points. The dashed blue lines enclose
  $95,5$\% of the data points at given Rosseland optical depth. The black
  solid line shows the <3D> model (the temperature of the 3D model averaged at
  constant $\tau_{\rm Ross}$), the black long-dashed line represents the
  $T(\tau_{\rm Ross})$ relation of the 1D LHD reference model atmosphere with
  identical stellar parameters, treating convection by standard mixing-length
  theory, but otherwise using the same input physics as the 3D hydrodynamical
  model.
}
\label{fig:3dt}
\end{figure}

As shown in Fig.\,\ref{fig:3dt},
both models are characterised by substantial photospheric temperature
fluctuations that cannot be captured by 1D (or 3D-averaged) models.\footnote{This is 
in stark contrast to the 3D temperature structure of
SDSS\,J102915.14+172927.9, one of the most metal-poor star known to date, where
horizontal temperature fluctuations are effectively suppressed
\citep[see][Figure\,2]{leo2024}.}
Spectral lines emerging from an inhomogeneous atmosphere can have a
significantly different strength (an shape) compared to the same lines formed
in a representative 1D (<3D>) atmosphere. In the present case, the
investigated CH, NH, OH molecular lines show a highly non-linear temperature
dependence. The strong temperature fluctuations in the 3D models lead to a
substantial enhancement of these lines, so that the abundances derived from
these models are significantly lower than those obtained from a standard 1D
analysis.

We use the line formation code
Linfor3D\footnote{\url{https://www.aip.de/en/members/matthias-steffen/linfor3d-user-manual/}}
to compute synthetic spectra from selected snapshots of the 3D hydrodynamical
model as well as from the 1D LHD reference model atmosphere with identical
stellar parameters and input physics as the 3D hydrodynamical model, except
that convection is treated by the mixing-length theory. Linfor3D treats
continuum scattering as true absorption which may compromise the synthetic
spectra in the UV, e.g., of the OH lines near $319,5$\,nm. However, the
abundance corrections are probably affected only marginally since the same
scattering approximation is used for the line formation in the 3D and the 1D
LHD models. 

3D abundance corrections are derived for individual spectral lines from a comparison of
3D and 1D equivalent widths. More specifically, if the observed equivalent width
of a spectral line of element X is reproduced with (logarithmic) abundance
$A_{\rm 3D}(X)$ based on the 3D model and with $A_{\rm 1D}(X)$ when using the
1D model, the 3D abundance correction is defined as $\Delta_{\rm 3D} = A_{\rm
3D}(X)-A_{\rm 1D}(X)$. For the molecular lines and the \ion{Fe}{II} line
investigated in this work, both 3D and 1D line formation are treated in LTE.

It must be noted that the 3D abundance correction of partly saturated lines
depends on the microturbulence velocity assumed in the 1D spectrum synthesis.
Here we have determined the optimal value of the microturbulence as $1.3$\,km\,s$^{-1}$ 
from a curve-of-growth analysis of the \ion{Fe}{II} line at $516.9$\,nm.

In the present work, the $^{12}$C/$^{13}$C isotopic ratio is derived from the
CH lines in the G-band where the $^{12}$CH and the $^{13}$CH lines are
well separated such that their individual equivalent widths can be measured. 
We can therefore compute individual 3D abundance corrections, $\Delta_{\rm
3D}^{12}$ and $\Delta_{\rm 3D}^{13}$ for both isotopic components as described
above. The isotopic ratio derived in 3D, $q_{\rm 3D}$, is related to the 1D
isotopic ratio, $q_{\rm 1D}$,  by $\log (q_{\rm 3D}/q_{\rm 1D}) =
\Delta_{\rm 3D}^{12 } - \Delta_{\rm 3D}^{13}$. If the corrections for both
components are identical, the 3D correction factor for the isotopic ratio is
$R_q = q_{\rm 3D}/q_{\rm 1D}=1$.

As mentioned in Sect.\,\ref{secabbo}, the magnitude of the 3D corrections
depends on the chemical composition, in particular the CNO abundances, adopted
for the construction of the 3D atmosphere and the 1D reference model.  This is
because the molecular opacities influence the resulting temperature structure,
as demonstrated in Fig.\,\ref{fig:3dt}. The model with the lower CNO
abundances (model A, top panel) shows a T-distribution that is strongly biased
towards the lowest temperatures, visible as the dark-orange band in the upper
photosphere, which leads to a plateau in the <3D> temperature profile around
$\log \tau_{\rm Ross}\sim -4$. In the model with enhanced CNO abundances
(model B, bottom), the T-distribution is more symmetric, and the <3D> and 1D
$T(\tau)$ relations are more similar throughout the photosphere.

The physical reasons for these differences need yet to be understood. 
For given carbon abundance, model A produces stronger CH lines due to the
excess of cool temperatures. It appears plausible that the larger asymmetry of
the T-distribution and the larger separation between <3D> and 1D reference
model lead to larger (more negative) 3D abundance corrections for the molecular
lines in model A than in model B. Since the 3D and 1D curves-of-growth are
running almost parallel in model B, the correction factor for the isotopic
ratio is close to $1$, while it is significantly below $1$ in model A.

Reliable 3D corrections can only be evaluated when a new 3D model is available
that is based on a (yet unavailable) custom-made opacity table representative
of the actual chemical composition of HE\,0107--5240.

\section{Ionised Fe in SMSS\,J160540.18--144323.1} \label{sec:nordlander}

\begin{figure}
\centering
\includegraphics[width=\hsize,clip=true]{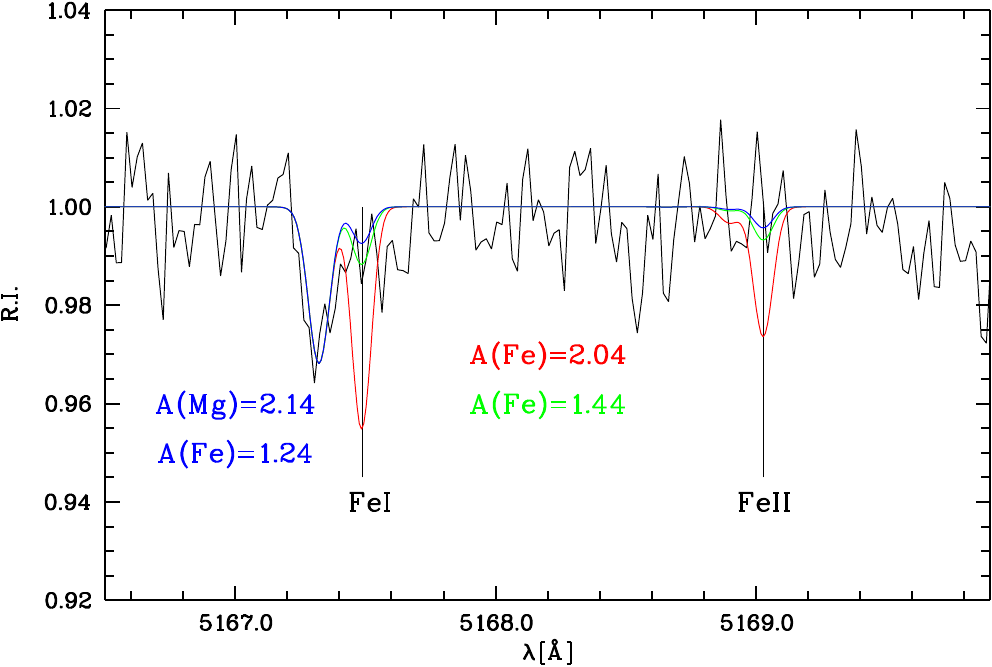}
\caption{Observed spectra (solid black) of SMSS\,J160540.18--144323.1, compared to 1D-LTE synthesis with different  A(Fe) values (solid blue, green and red).}
\label{fig:nordlanderfe}
\end{figure}

The problem, discussed in Sec.\,\ref{sec:discussfe2}, about the large NLTE corrections for \ncstar\ proposed in the literature for A(Fe) from \ion{Fe}{i} lines \citep[see][]{ezzeddine17,nordlander2017} and that bring tension with A(Fe) from \ion{Fe}{ii} lines, is also true for SMSS\,J160540.18--144323.1 \citep{nordlander2019}.
We took the advantage of 18 ESPRESSO observations to investigate the \ion{Fe}{ii} line at 516.9\,nm, the same one investigated in \ncstar\ and shown in the lower panel of Fig.\,\ref{fig:fe2}.
We here suppose that the NLTE correction for \ion{Fe}{ii} line is negligible for this star as for \ncstar.
In Fig.\,\ref{fig:nordlanderfe} the observed spectrum of SMSS\,J160540.18--144323.1 (solid black) is compared to a synthesis (solid blue) with the LTE iron abundance  derived from neutral Fe lines  \citep[A(Fe)=1.24][]{nordlander2019}: the agreement observation/synthesis is good for both the \ion{Fe}{i} and the \ion{Fe}{ii} line, which is not detected.
A NLTE correction of +0.2\,dex in \ion{Fe}{i} (a value compatible with the NLTE correction for a more metal rich model derived in the MPIA database and discussed in Sec.\,\ref{secabbo}) would bring A(Fe)=1.44: the solid green synthesis in the figure shows that the profile is compatible with the observations of the \ion{Fe}{ii} line.
The `large' NLTE correction on \ion{Fe}{i} \citep[as suggested by][]{nordlander2017,ezzeddine17} is not compatible with the observation. A NLTE correction of $+0.8$\,dex would bring the LTE A(Fe) value at 2.04, and this is shown as a red solid profile in the figure. Clearly such a NLTE correction would imply the \ion{Fe}{ii} line to be visible at this observation quality, and this is not the case.
A dedicated work will perform a careful investigation of this line as well as a complete analysis of this ESPRESSO spectrum.

\end{appendix}

\end{document}